\pdfoutput=1

\documentclass[11pt]{article}

\usepackage
{acl}

\usepackage{times}
\usepackage{latexsym}
\usepackage{graphicx}
\usepackage{todonotes}
\usepackage{multirow}
\usepackage{subcaption}
\usepackage{booktabs}

\usepackage[T1]{fontenc}

\usepackage[utf8]{inputenc}

\usepackage{microtype}

\usepackage{inconsolata}

\title{Social Norms in Cinema: A Cross-Cultural Analysis of \textit{Shame}, \textit{Pride} and \textit{Prejudice} }

\author{Sunny Rai$^\dag$, Khushang Jilesh Zaveri$^\diamond$, Shreya Havaldar$^\dag$, Soumna Nema$^\spadesuit$, \\ \bf Lyle H. Ungar$^\dag$ \& Sharath Chandra Guntuku$^\dag$   \\
$^\dag$University of Pennsylvania, $^\diamond$Johns Hopkins University, $^\spadesuit$Mahindra University \\
\texttt{\{sunnyrai, sharathg\}@seas.upenn.edu} \\}

\begin{document}
\maketitle

\begin{abstract}
Shame and pride are social emotions expressed across cultures to motivate and regulate people's thoughts, feelings, and behaviors.  In this paper, we introduce the first cross-cultural dataset of over $10k$ shame/pride-related expressions, with underlying social expectations from ~5.4K Bollywood and Hollywood movies. We examine \textit{how} and \textit{why} {shame and pride} are expressed across cultures using a blend of psychology-informed language analysis combined with large language models. We find significant cross-cultural differences in shame and pride expression aligning with known cultural tendencies of the USA and India -- e.g., in Hollywood, shame-expressions predominantly discuss \textit{self} whereas shame is expressed toward \textit{others} in Bollywood. Women are more sanctioned across cultures and for violating similar social expectations.

\end{abstract}

\section{Introduction} 
Shame and pride are social emotions expressed across cultures to motivate and regulate people's thoughts, feelings, behaviors, achievements, and social adaptations \cite{fischer1995self,goetz2007shifting, fessler2007appeasement, schaumberg2022shame}. The expressions and appraisals of shame and pride vary across cultures and is a research question of great significance to understand society and self-regulation \cite{wong2007cultural, lewis2010cultural, wegge2022experiencer}. 
 The overarching question we ask in this paper is,  \textbf{\textit{how} and \textit{why} shame and pride are expressed across cultures}. Knowing cross-cultural differences in shame/pride-related expressions will help us measure psychological constructs such as shame proneness \cite{tangney1992proneness, cohen2011introducing} and self-esteem \cite{pyszczynski2004people} reliably from language -- enabling effective human-computer interactions in domains such as AI-driven psychotherapy \cite{resendiz2023affective, demszky2020goemotions}.
Examining behaviors evoking {shame} and {pride} will reveal culture-specific societal expectations and attitudes (See Table \ref{tab:dialogue_example}) -- paramount to building culturally competent AI models for diverse users \cite{talat2021word,atari2023humans}. 

\begin{table}[t]
    \centering
    \begin{tabular}{p{7.5cm}}
  \\ \toprule
  \small
\texttt{And should we bow before others begging....them to marry our daughters? This shall not happen. Neither will the girls be alive here nor shall....\textbf{we be ashamed of ourselves}. 
You cannot kill the life which God has given. I won't let you commit the sin. }
\\ 

  $\Rightarrow$ \textcolor{red}{\textbf{Not able to marry off their daughters} evokes shame.} \\ \midrule
 \small
   \texttt{Sister-in-law! Congrats, sister-in-law! Big brother has started working! \newline
   Really? \newline
\textbf{Now you will have a place of pride in this family}! \newline
   Yes, please! 
 } \\ 
$\Rightarrow$ \textcolor{blue}{\textbf{Employed husband} evokes \textit{pride}} \\\bottomrule
         
    \end{tabular}
    \caption{Excerpts of dialogues expressing explicit social emotions shame and pride indicating actions and social approval. }
    \label{tab:dialogue_example}
\end{table}


For our study, we select two top movie industries, i.e., Hollywood and Bollywood\footnote{\url{https://en.wikipedia.org/wiki/Film_industry}}.  Hollywood primarily depicts social situations from the USA, an individualist society that values competency and autonomy \cite{triandis1989self, triandis1988collectivism}. Bollywood depicts India, a collectivist society where one's sense of self is interwoven with community beliefs. The cultural dichotomy between India and the USA (i.e., \textit{collectivism vs. individualism}) thus presents a rich ground for understanding variations in their beliefs and values. Our approach blends psychology-informed language analysis with state-of-the-art Large Language Model (LLM) to delineate interpretable psychosocial constructs such as \textit{self-focus} and \textit{morality} in shame/pride-related expressions and extract implicit social expectations behind them.

This paper makes the following research contributions: (a) We develop a cross-cultural dataset of over 10k shame/pride-related dialogues with underlying social expectations\footnote{\url{https://github.com/Khushangz/Cross-Cultural-Social-Norms-Dataset/}}, (b) We demonstrate cross-cultural variations in how and why shame and pride are expressed in movies, and
(c) We illustrate that women are more subjected to social sanctions than men across cultures and for violating similar social expectations.

 To the best of our knowledge, this is the first empirical study analyzing the psychosocial constructs underlying shame/pride-related expressions across cultures and the social expectations behind them.



\section{Movies Subtitles Corpora}
 Movies provide culture-specific 
life-like depictions of social situations \cite{adkins2014moving,kubrak2020impact}. The natural conversation style between characters in movies can reveal the social power dynamics (e.g., \textit{boss-employee}, \textit{father-daughter}) and gender roles.
Another potential alternative is social media posts however, they are (a) skewed to social situations specific to young demographic groups from Western countries, and (b) are second-hand reports that do not reveal fully the social dynamics of how shame/pride are expressed in natural conversations, for instance, between a boss and an employee. 

We collected English subtitles for 5,435 Hollywood and Bollywood movies that were released post-1990 by auto-crawling
websites that host or link movie subtitles (See Table \ref{tab:data_stats} for data distribution). 
The subtitles are professionally done translations to preserve the emotions and intended message.  Bollywood movie subtitles were collected from \url{www.Bollynook.com} and Hollywood movie subtitles were from the publicly available Kaggle movie subtitles dataset\footnote{\url{https://www.kaggle.com/datasets/adiamaan/movie-subtitle-dataset}}. The year mapping was performed to ensure a similar period for collected movies. 
The year of release for movies was verified by either parsing subtitle file names having a release year or probing Wikipedia entries.

\begin{table}[t]
    \centering
    \begin{tabular}{l|l|l}
    \toprule
 &  \textbf{Hollywood} & \textbf{Bollywood}\\ \midrule
\#Movies & 2697  &  2738 \\ 
\#Tokens &  20.78M & 22.62M  \\ 
  \#Shame  &  1221 &   5409 \\
  \#Pride & 2805 & 2999 \\
  \#Control & 4385 & 8303 \\  \bottomrule
    \end{tabular}
    \caption{Data Distribution. \#shame indicates the number of dialogues with the word "shame" or its derivative form (e.g., ashamed, shameless - See Table \ref{tab:search_keywords}). Similarly, \#pride indicates the number of dialogues with the word pride or its derivative form (e.g., proud). Control comprises dialogues without words \textit{shame, pride}, or their derivatives. }
    \label{tab:data_stats}
\end{table}

\subsection{Extracting Shame and Pride related Expressions} 
The expressions of shame and pride could be explicit (See Table \ref{tab:dialogue_example}) or implicit (e.g. characters' lowered shoulders, avoiding eye contact, etc.). We extracted \textbf{explicit mentions of \textit{shame} and \textit{pride}} to learn about occurrences of shame and pride \textit{as determined by the characters}. We adopted a keyword search-based approach (See Appendix \ref{tab:search_keywords}) to identify dialogues along with the previous and the next two lines for situational context (See Table \ref{tab:dialogue_example}). For short dialogues such as monosyllabic responses in spoken conversations, we appended an extra previous and next line to context. One author manually checked the entire dataset of shame and pride-related dialogues to 
filter out dialogues with conventional phrases such as "\textit{what a shame}", "\textit{it's a shame}", "\textit{proudly presents}".

Detecting \textbf{implicit shame/pride-related expressions} in textual discourse is a complex task. Computational systems for detecting social emotions such as shame and guilt underperform significantly compared to basic emotions such as joy due to internal self-evaluation associated with social emotions \cite{demszky2020goemotions, resendiz2023affective}. We used GPT-4 language model \cite{achiam2023gpt} to detect implicit expressions of shame in given movie dialogues. We recruited two human annotators to determine the agreement with GPT-4 generated labels (See Appendix \ref{sec:implicit_shame} for details). On a randomly sampled 100 situations each from H/Bollywood, the model labeled 20 samples to be expressing shame in Bollywood and 5 samples in Hollywood. However, the agreement between human annotators was low (4 out of 20 samples in Bollywood and 1 out of 5 samples in Hollywood where both annotators agreed with GPT-4). 
Human annotators across cultures tend to cross-label shame and guilt \cite{troiano2019crowdsourcing} and follow-up discussions with annotators revealed similar tendencies. One such example from Hollywood is provided below:

``\textit{about feeding the poor, I've never done any of that. 
 God tells us to love everybody. 
 I've hated people... my family,  my family,} ''

Here, the speaker is probably expressing guilt, shame, or both. Given subjective beliefs and poor agreement with generated outputs, we adopted a high precision-low recall strategy and limited our further analysis to explicit mentions of shame and pride.

We created four sets of dialogues: (a) shame-related dialogues in Bollywood, (b) shame-related dialogues in Hollywood, (c) pride-related dialogues in Bollywood and (d) pride-related dialogues in Hollywood. Cross-cultural differences also exist in speaking styles across cultures and we thus formed a control set of dialogues unrelated to \textit{shame} and \textit{pride} for both movie industries to remove the variations in language markers owing to culture-specific speaking styles (See Table \ref{tab:data_stats} for dialogue distribution).

 \begin{figure}[t]
    \centering
    \includegraphics[width=\linewidth]{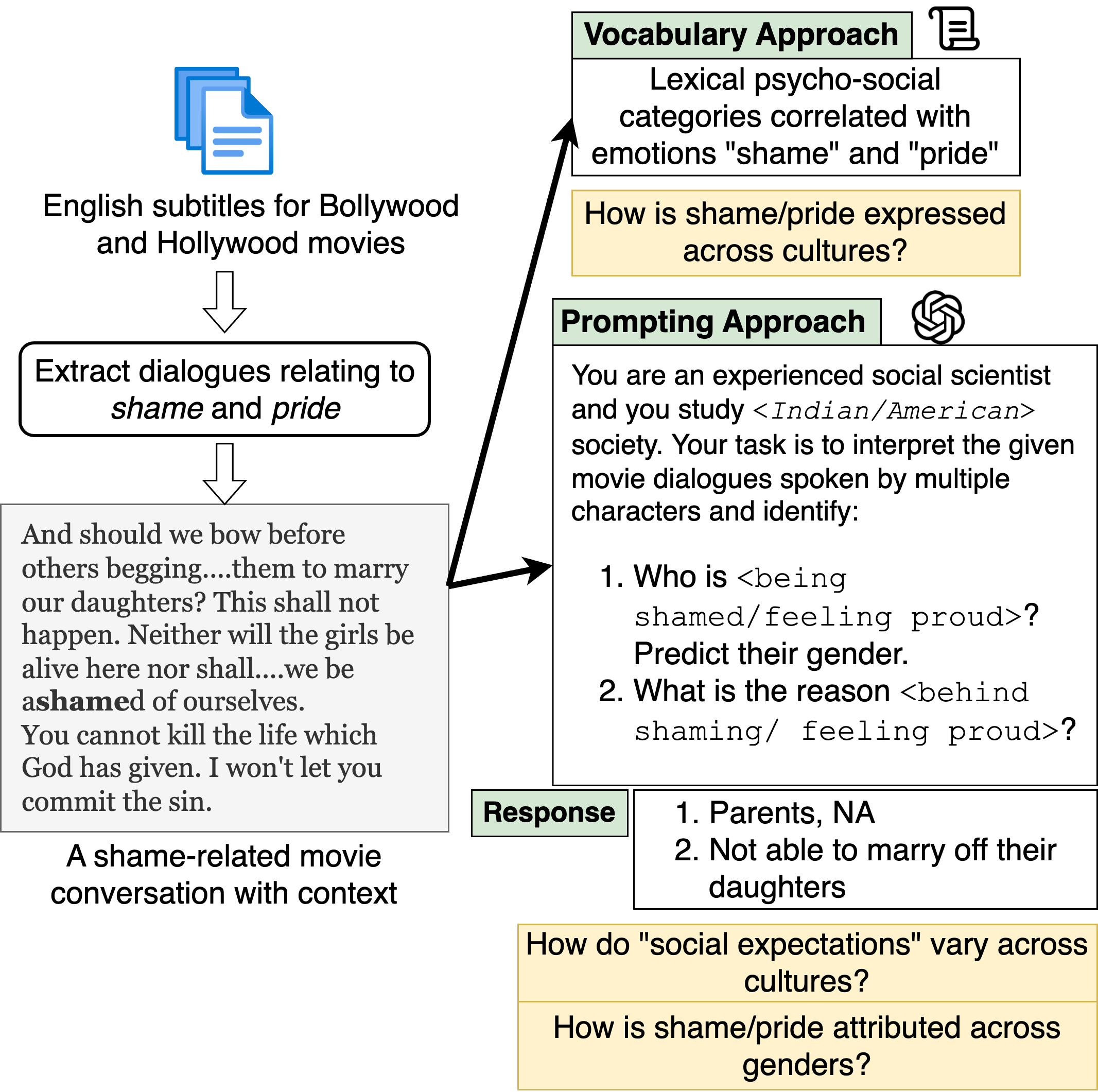}
    \caption{An overview of our approach comprising two key steps (a) Vocabulary approach and (b) Prompting a pre-trained LLM.}
    \label{fig:approach}
\end{figure}

\section{Approach}
Fig. \ref{fig:approach} illustrates our study design comprising two key approaches (a) using a psychosocial vocabulary approach to measure how expressions of shame and pride differ and (b) \textit{prompting LLM} approach to extract reasons behind them and how they differ between India and the U.S.A. 

\subsection{Vocabulary Approach}
Linguistic Inquiry of Word Count (LIWC) \cite{boyd2022development} is a corpus analysis tool to identify
psychological constructs such as "self-focus" and "morality", and therefore, is widely used for examining social behaviors such as self-regulation and conformity. 

To understand {cross-cultural linguistic variations in the manifestation of shame and pride}, we computed the normalized distribution of psychosocial categories in LIWC from the dialogues and examined their correlation with shame and pride compared to the control set.  
{The search keywords used for building social emotions corpus (See Table \ref{tab:search_keywords}) were removed from the LIWC dictionary to prevent overestimation of shame- (e.g., negative emotion) and pride-related categories (e.g., achievement).}

    \subsection{Prompting Approach}
Identifying why shame/pride is expressed in discourse requires multicultural world knowledge. LLMs such as GPT-4 exhibit a superlative pragmatic understanding of the world around us and are increasingly used for extracting implicit meanings and beliefs \cite{pan2023rewards,tornberg2023chatgpt}. We used GPT-4 chat in a two-shot setting (
See Tables \ref{tab:prompts_bolly} and \ref{tab:prompts_holly} for prompts) to 
identify: 
\begin{itemize}
    \item who is $<$being shamed/feeling proud$>$ in the given movie discourse, and what is their gender? and,
    \item What is the reason behind $<$the feeling of shame/pride$>$? 
\end{itemize}

Since there are at least two characters in a discourse, {the first question orients the LLM to focus on the \textit{person experiencing the social sanction or approval} and then identify their gender. The output for the second question serves as the implicit \textit{social expectation} that led to the expression of shame/pride in the culture.}  Asking for "reason" leading to the expression of social emotion encourages LLMs to retrieve expectations from the provided context, mitigating potential Anglocentric tendencies \cite{havaldarEmotion}.
 
\paragraph{Thematic Analysis of Reasons Behind Shame/Pride-related Expressions}
 To capture overarching themes in social expectations that led to shame/pride-related expressions in Indian and American societies, we performed agglomerative clustering after embedding unique shame and pride-related reasons using SBERT embeddings \cite{reimers2019sentence}. 
Unlike movie dialogues, reasons extracted from GPT-4 chat are short phrases (See Table \ref{tab:reason_shame}) and are devoid of culture-specific language style markers (i.e., Indian English vs American English) thus, no control for culture-specific language is needed.  

\begin{figure}[t]
    \centering
    \includegraphics[width=1\linewidth]{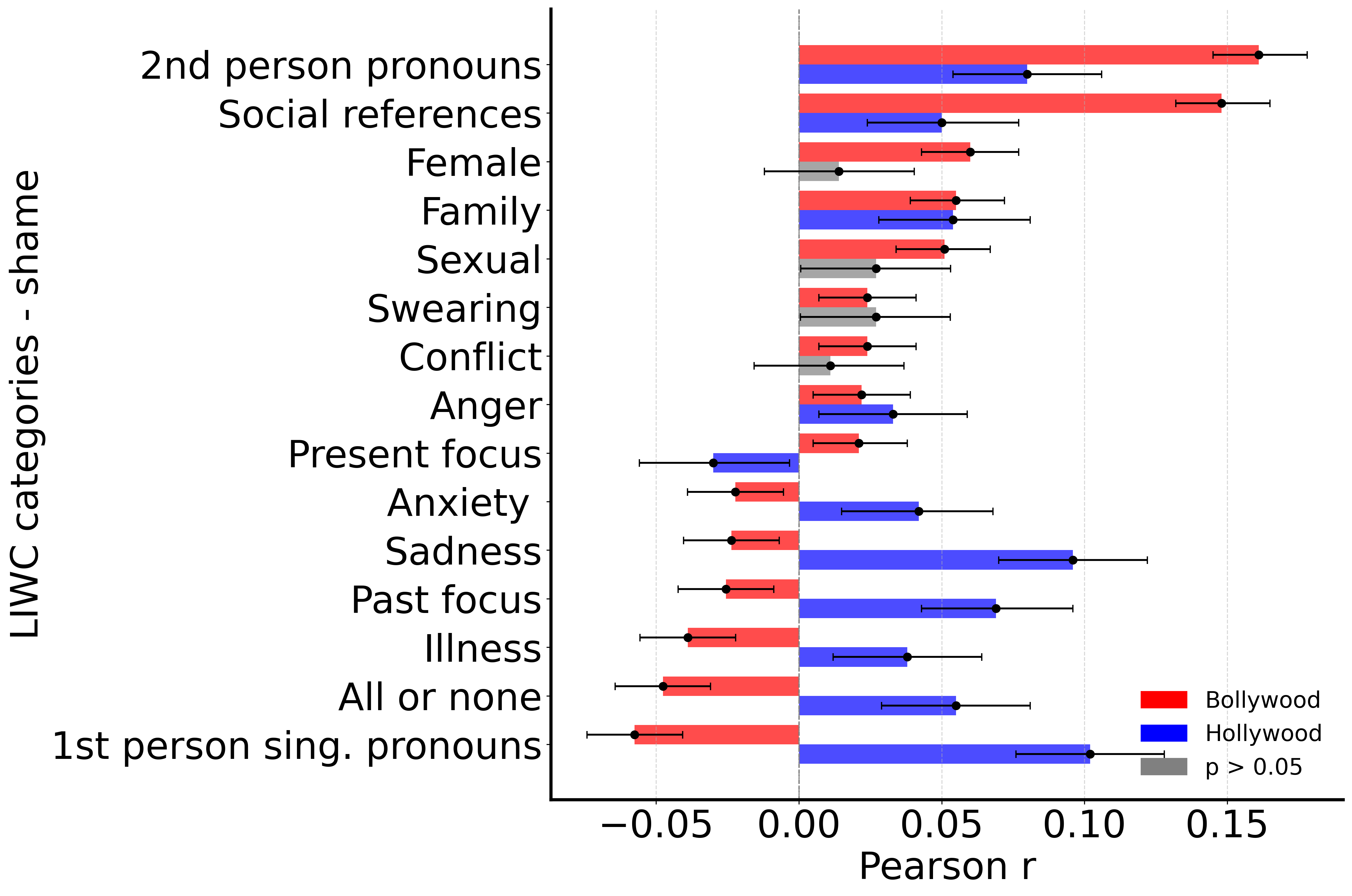}
    \caption{Pearson \textit{r} for LIWC categories significantly correlated ($p<0.05$) with \textbf{shame} for Hollywood and Bollywood after Benjamini-Hochberg p-value correction (See Appendix for confidence intervals and p values). Note, $1^{st}$  person sing pronouns are strongly correlated with Hollywood-shame whereas the correlation with $2^{nd}$ person pronoun and social references are up to 3 times stronger in Bollywood compared to Hollywood.  See Table \ref{tab:liwc_shame_bolly} and \ref{tab:liwc_shame_holly} for the complete set of correlations.}
    \label{fig:shame_compare}
\end{figure}

\begin{figure}[t]
    \centering
    \includegraphics[width=1\linewidth]{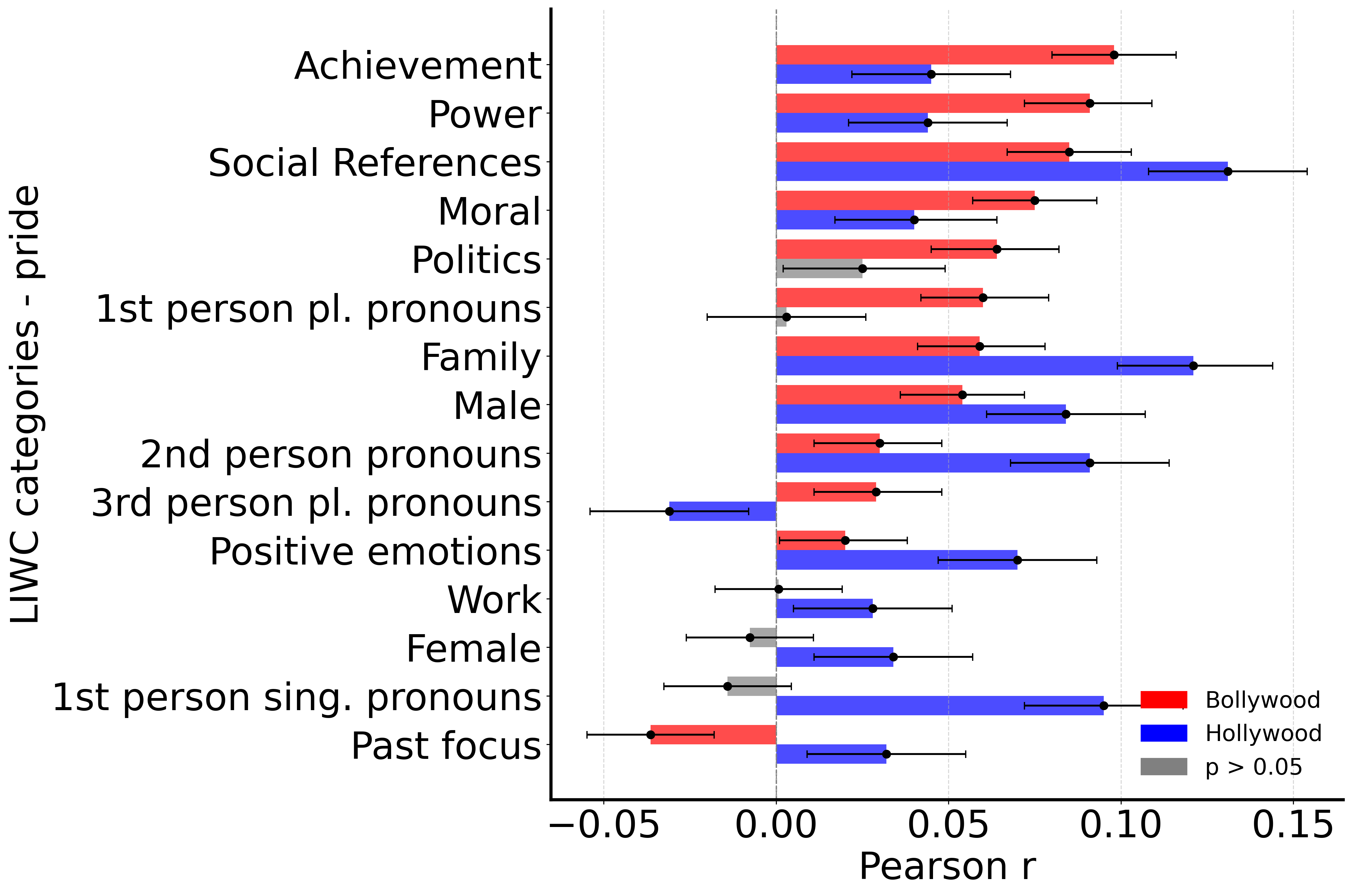}
    \caption{Pearson \textit{r} for LIWC categories significantly correlated ($p<0.05$) with \textbf{pride} for Hollywood and Bollywood after Benjamini-Hochberg p-value correction (See Appendix for confidence intervals and p values). Note the contrast, Achievement-related \& We-centered pride in Bollywood vs Social \& Self-centered pride in Hollywood.  See Table \ref{tab:liwc_pride_bolly} and \ref{tab:liwc_pride_holly} for the complete set of correlations.}
    \label{fig:pride_compare}
\end{figure}

\section{Results}
\subsection{ How is shame/pride expressed across cultures? } \label{sec:liwc_results}


Shame is associated with negative emotions, power, and morality in both movie industries (See Tables \ref{tab:liwc_shame_bolly} and \ref{tab:liwc_shame_holly}) however, significant cultural nuances exist in its manifestation (See Fig. \ref{fig:shame_compare}).  In Bollywood, shame-related expressions are \textit{other-oriented} as indicated by 2nd person pronouns and social references. Female and sexual categories are correlated with shame exclusively in Bollywood potentially indicating the honor system in collectivist societies \cite{caffaro2014gender}. Other psychosocial categories such as \textit{swearing} and \textit{conflict} with present focus reveal the role of shame in enforcing conformity. In Hollywood, shame expressions are self-focused with remorse (e.g., \textit{sadness, anxiety}) and have past-focused language. \textit{Illness} is uniquely correlated with shame in Hollywood indicating social sanctions around incapability.  
 
  A similar dichotomy is observed for {pride} (See Fig. \ref{fig:pride_compare}). In Bollywood, pride is \textit{achievement/power} focused and "We"-centered. Pride is associated exclusively with men in Bollywood. Other psychosocial categories such as \textit{moral} and \textit{politics} are more strongly associated with pride. In Hollywood, pride is more self-centered and expressed for \textit{family} more. Male references are more strongly associated with pride than females in Hollywood whereas the female category has an insignificant ($p>0.05$) correlation with pride in Bollywood. Pride has a positive undertone in Hollywood whereas in Bollywood, we speculate that it relates to honor (e.g., \textit{protecting family's pride, bringing pride to family}) and does not have a positive undertone.


\subsection{How do \textit{social expectations} behind shame/pride vary across cultures? } 

\paragraph{Cross-Cultural Shame/Pride Dataset} For the Bollywood set, GPT-4 predicted reasons (also implicit social expectations) for 5321 (98.4\%) shame-related dialogues out of 5409, and 2237 (74.6\%) pride-related dialogues out of 2999. For the Hollywood set, GPT-4 chat predicted a reason for 1156 (94.6\%) shame-related dialogues out of 1221 and 1731 (61.7\%) pride-related dialogues out of 2805. Upon manual analysis, we found that pride is also used to express affection, specifically toward close family members such as children. As a result, a lower number of reasons is associated with pride compared to shame. GPT-4 chat predicted a total of \textbf{10,445 social expectations} (See Table \ref{tab:normGenderDistribution} for distribution). Prompting "reasons" behind shame and pride allowed us to capture high specificity in cultural norms (See Tables \ref{tab:reason_shame} for shame-related expectations and \ref{tab:reason_pride} for pride-related expectations).

\subsubsection{Manual Evaluation} Two volunteers manually verified the predicted gender for "the person experiencing shame/pride" and the predicted "reason"  for a randomly sampled set of $100$ dialogues each from Hollywood and Bollywood. 
An Indian annotator aware of social roles and expectations in Indian society labeled the Bollywood set. Likewise, an American annotator labeled the Hollywood set.

For Bollywood, 8\% of predicted gender and 11\% of predicted reasons were incorrect where it was 5\% and 2\% respectively for Hollywood (See Table \ref{tab:manual_evaluation} for more details). There were 20 cases where gender (15) and/or reason (9) were ambiguous whereas, for Hollywood, there were 11 (gender=10, reason=1) such samples. 

\begin{table}[t]
    \centering
    \begin{tabular}{c|c |c |c|c}
   
      &  \multicolumn{2}{c|}{Bollywood} & \multicolumn{2}{c}{Hollywood}\\ \toprule
     & shame & pride & shame & pride \\ \midrule
       male & 2690 & 1259 & 591& 776 \\
       female & 1215 & 326 & 246 & 236 \\ 
       
       Not clear & 1416 & 652 & 319&  719\\ \midrule 
       total & 5321 & 2237 & 1156 & 1731 \\ \toprule
       
    \end{tabular}
    \caption{Reason and Gender distribution (with duplicates) for dialogues for which GPT-4 predicted \textit{male} or \textit{female}. The duplicate reasons are not removed as their frequency reflects their prevalence and is useful for estimating gender association. }
    \label{tab:normGenderDistribution}
\end{table}

\subsubsection{Thematic Analysis}  
\begin{figure}[t]
    \centering
    \includegraphics[width=\linewidth]{
    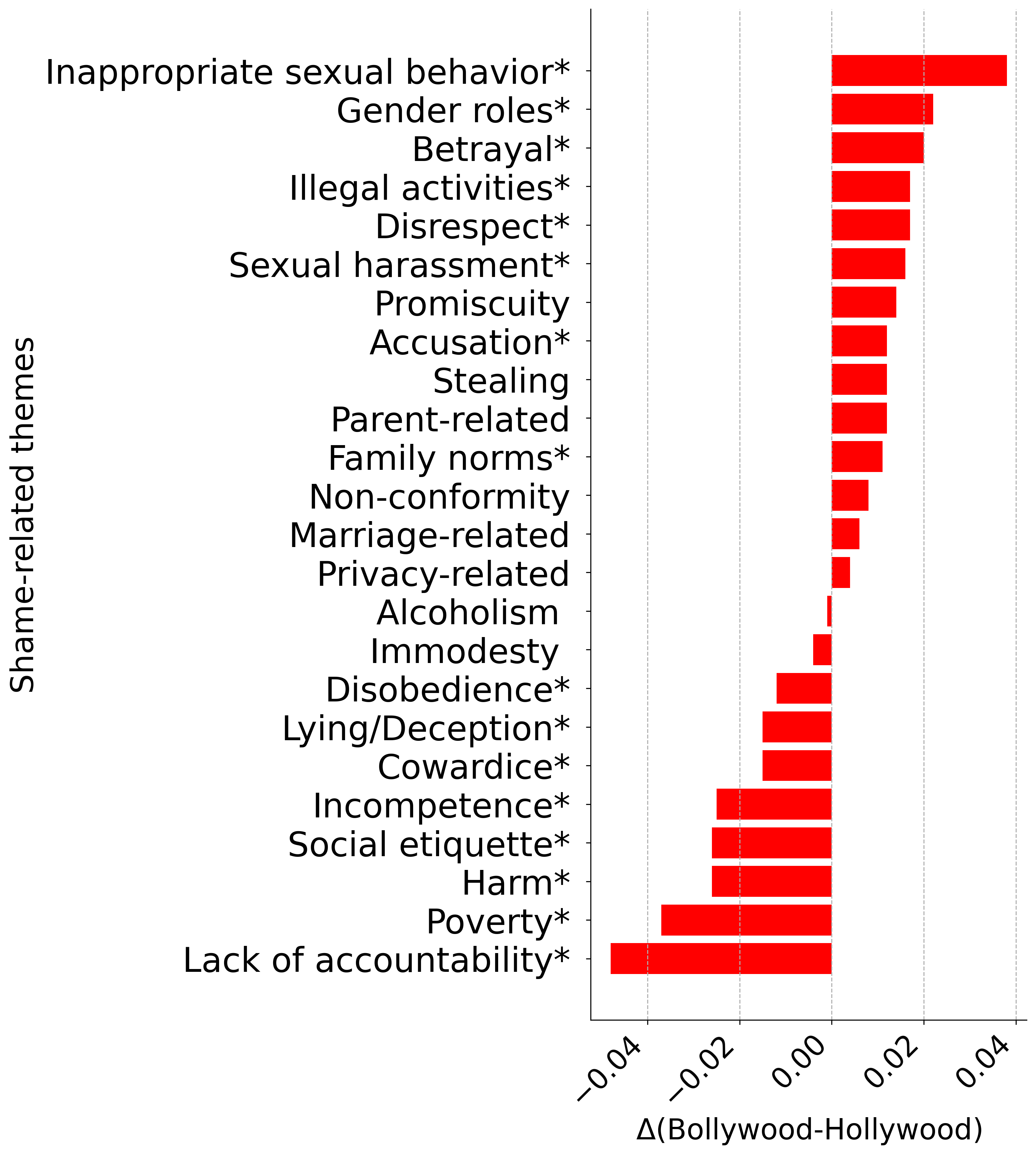}
    \caption{Relative association ($\Delta$) of Bollywood and Hollywood to themes obtained from agglomerative clustering performed on shame-related norms. $*$ indicates significant difference i.e., $p<0.05$. See Table \ref{tab:shame_statTest} for statistics and p-value.}
    \label{fig:cluster_shame}
\end{figure}

\begin{figure}[!tp]
\centering
    \includegraphics[width=\linewidth]{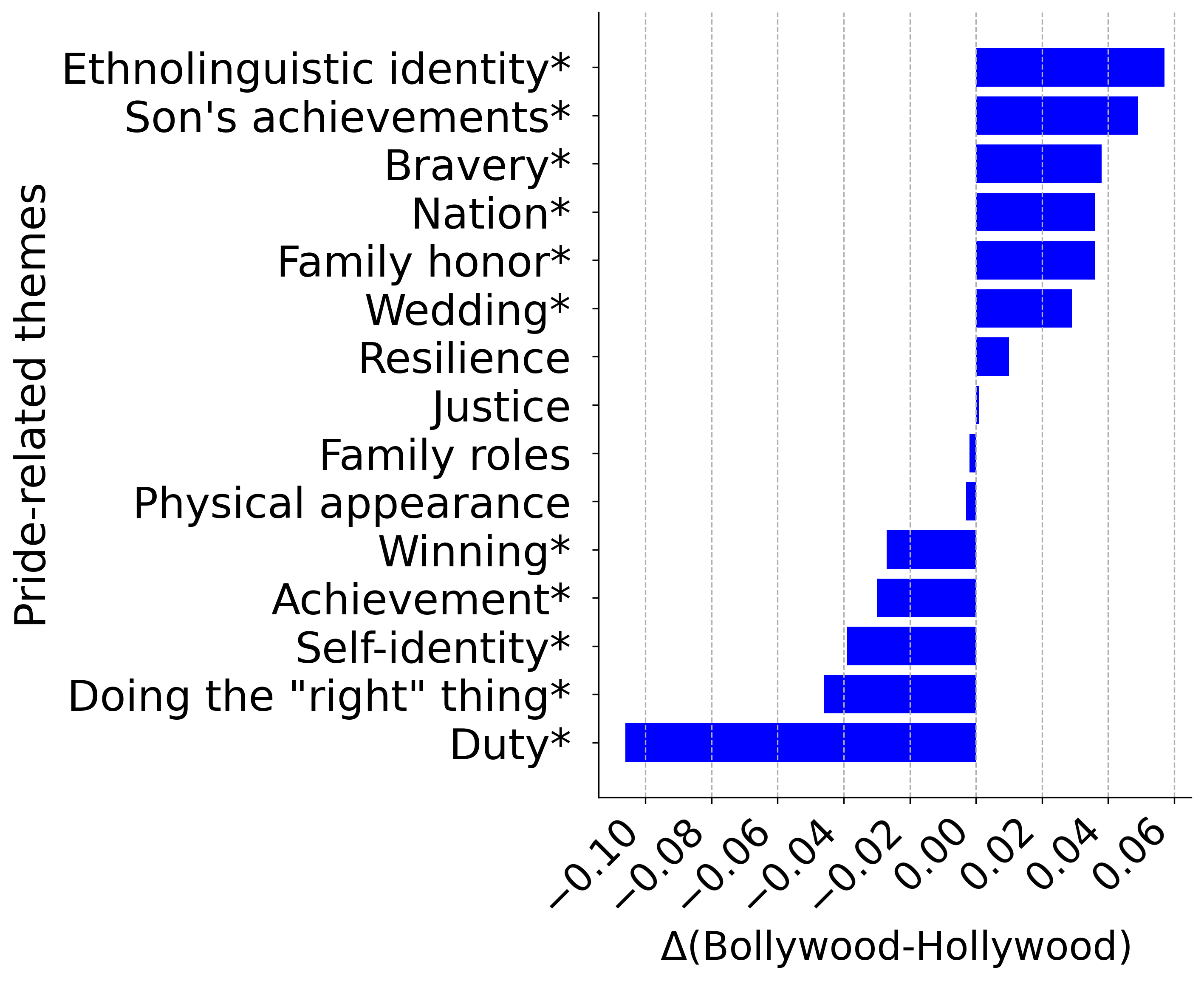}
    \caption{Relative association ($\Delta$) of Bollywood and Hollywood to themes obtained from agglomerative clustering performed on pride-related norms. $*$ indicates significant difference i.e., $p<0.05$. See Table \ref{tab:pride_statTest} for statistics and p-value.}
    \label{fig:cluster_pride}
\end{figure}

Twenty-four clusters for shame-related reasons and fifteen for pride-related reasons were formed using agglomerative clustering ( See Tables \ref{tab:clusterThemes_shame} and \ref{tab:clusterThemes_pride} for clustering parameters). The clusters were manually assigned a theme (as depicted on the Y-axis of Fig. \ref{fig:cluster_shame} and \ref{fig:cluster_pride}) after analyzing the ten closest samples based on cosine distance from the centroid of the cluster (See Table \ref{tab:shame_clusterEx} and \ref{tab:pride_clusterEx} for examples in each cluster).  We computed the relative association for each cluster with Bollywood and Hollywood using eq. \ref{eq:cluster_theme} and performed the Barnard-Exact Test \cite{barnard1947significance} with the Yates Correction \cite{yates1934contingency} to test if the possibility of observing norms related to pre-assigned themes is statistically different across movie industries. 

\begin{table}[t]
    \centering

    \begin{tabular}{p{6.5 cm}}
     \toprule
       \textbf{Bollywood}  \\ \midrule
eavesdropping on private conversation \\
expressing love for a man \\
Incestuous relationship \\
giving birth to a girl child \\
\midrule
\textbf{Hollywood} \\ \midrule
not living up to expectations \\
hiding/avoiding confrontation\\
 not returning calls after intimacy \\
 mistreatment of a woman \\
\bottomrule

    \end{tabular}
    \caption{A subset of reasons extracted from movie dialogues expressing shame. A total of 4604 unique reasons (Bollywood-3660, Hollywood-944) were extracted.}
    \label{tab:reason_shame}
\end{table}

\begin{equation}
  \vec\Delta   = \forall_{t_i\in themes} \frac{\mathcal{D}_{bolly_{t_i}}}{{\mathcal{D}_{bolly}}} - \frac{\mathcal{D}_{holly_{t_i}}}{\mathcal{D}_{holly}}
  \label{eq:cluster_theme}
\end{equation}

Here, $\mathcal{D}_{industry_{t_i}}$ represents dialogues to $industry \in \{holly, bolly\}$ and ${t_i} \in themes$ (as depicted on Y-axis in Fig \ref{fig:cluster_shame} and \ref{fig:cluster_pride}).

 \paragraph{Shame-related social norms} 
Themes such as \textit{lack of accountability} and \textit{poverty} are more common in Hollywood whereas \textit{inappropriate sexual behavior} and \textit{gender roles} are more prevalent in Bollywood (See Fig. \ref{fig:cluster_shame}). Unsurprisingly, collectivist factors such as non-conformity in \textit{gender} roles, disrespect, and deviation from \textit{family norms} are strongly associated with shame in Bollywood whereas individualistic attributes such as \textit{poverty}, causing \textit{harm} and \textit{incompetence} evoke shame in Hollywood. 
 
 \paragraph{Pride-related social norms} \textit{Duty}, \textit{doing the "right" thing}, and, \textit{self-identity} are associated with pride in Hollywood whereas \textit{Ethnolinguistic identity}, and \textit{son's achievements} are associated with pride in Bollywood (See Fig. \ref{fig:cluster_pride}). 

\subsubsection{How is shame/pride attributed across genders?}  For the Bollywood set, GPT-4 predicted 1541 targets as female and 3949 as male. For the Hollywood set, GPT-4 predicted 482 targets as female and 1367 as male. Across all combinations (shame vs pride x Bollywood vs. Hollywood in Table \ref{tab:normGenderDistribution}), there are more male targets than females in line with the findings of Geena Davis Inclusion Quotient\footnote{\url{https://about.google/intl/ALL_us/main/gender-equality-films/}}. 

We computed the gender-wise attribution to "shame" and "pride" in movie dialogues using eq. \ref{eq:emotion_genderassociation}.  A positive score indicates a higher association of gender groups with pride, whereas a negative score reflects a higher association with shame. A null score indicates no preference. 

\begin{figure}[ht]
    \centering
    \includegraphics[width=\linewidth]{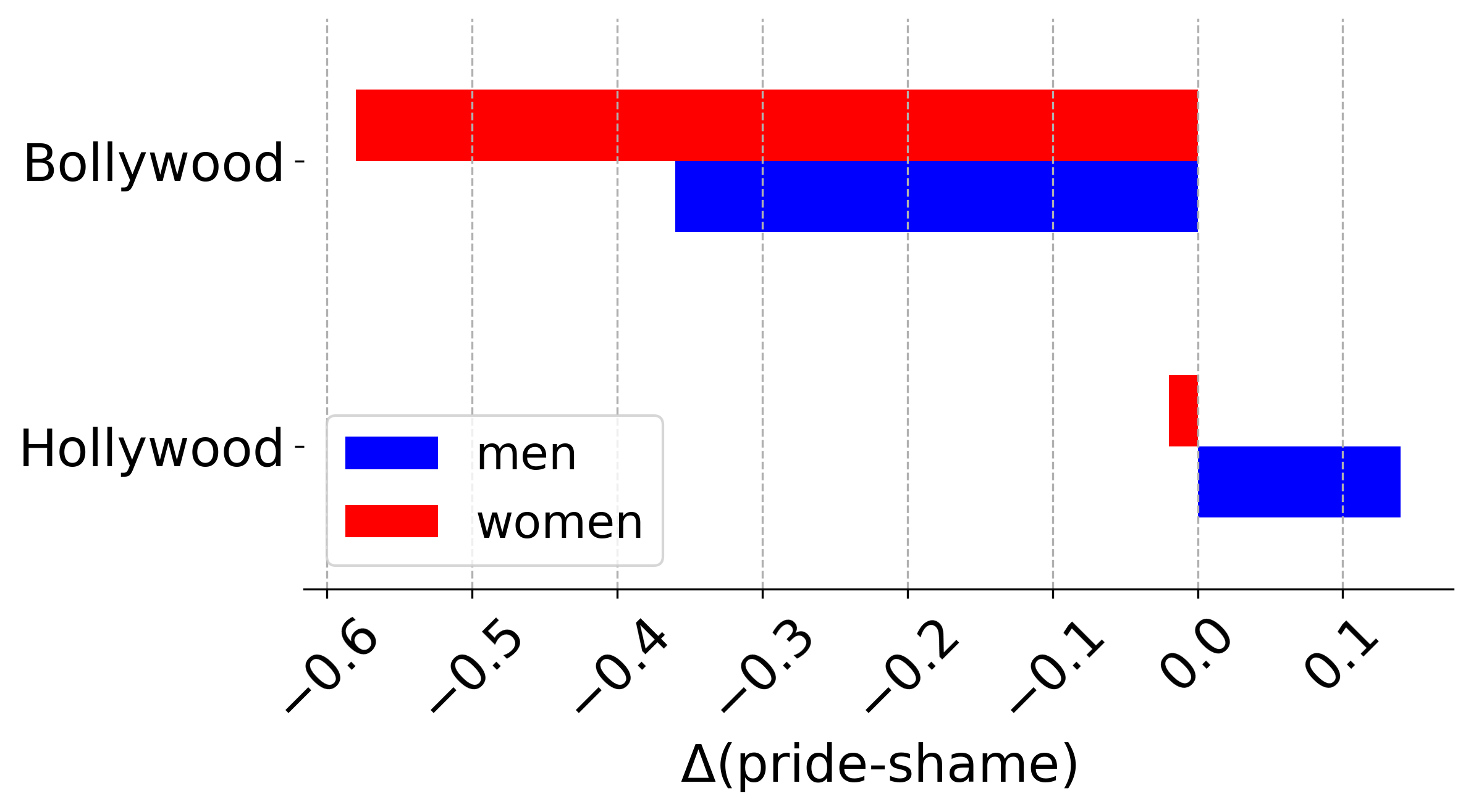}
    \caption{Relative association ($\Delta$) of social emotions attributed to male and female. A higher positive score indicates a stronger
association of gender with pride.}
    \label{fig:emotionwise_genderassociation}
\end{figure}

\begin{equation}
 \vec\Delta_g = \forall_{g \in \{male,female\}} \frac{\mathcal{D}_{pride_{g}}-\mathcal{D}_{shame_{g}}}{\mathcal{D}_{g}}
  \label{eq:emotion_genderassociation}
\end{equation}

Here, $\mathcal{D}_{pride_{g}}$ represents movie dialogues having pride-related words where the target of pride is $g\in\{male, female\}$,  $\mathcal{D}_{shame_{g}}$ represents movie dialogues having shame-related word where the target of shame is $g\in\{male, female\}$ and $\mathcal{D}_{g}$ represents the number of movie dialogues spoken by $g$.
As depicted in Fig. \ref{fig:emotionwise_genderassociation}, Hollywood movies are pride-oriented, whereas Bollywood movies are shame-oriented. Females are attributed more shame, and the difference (male-female) in the expression of pride and shame is $0.16$ for Hollywood and $0.21$ for Bollywood.

\begin{figure}[t]
    \centering
    \includegraphics[width=\linewidth]{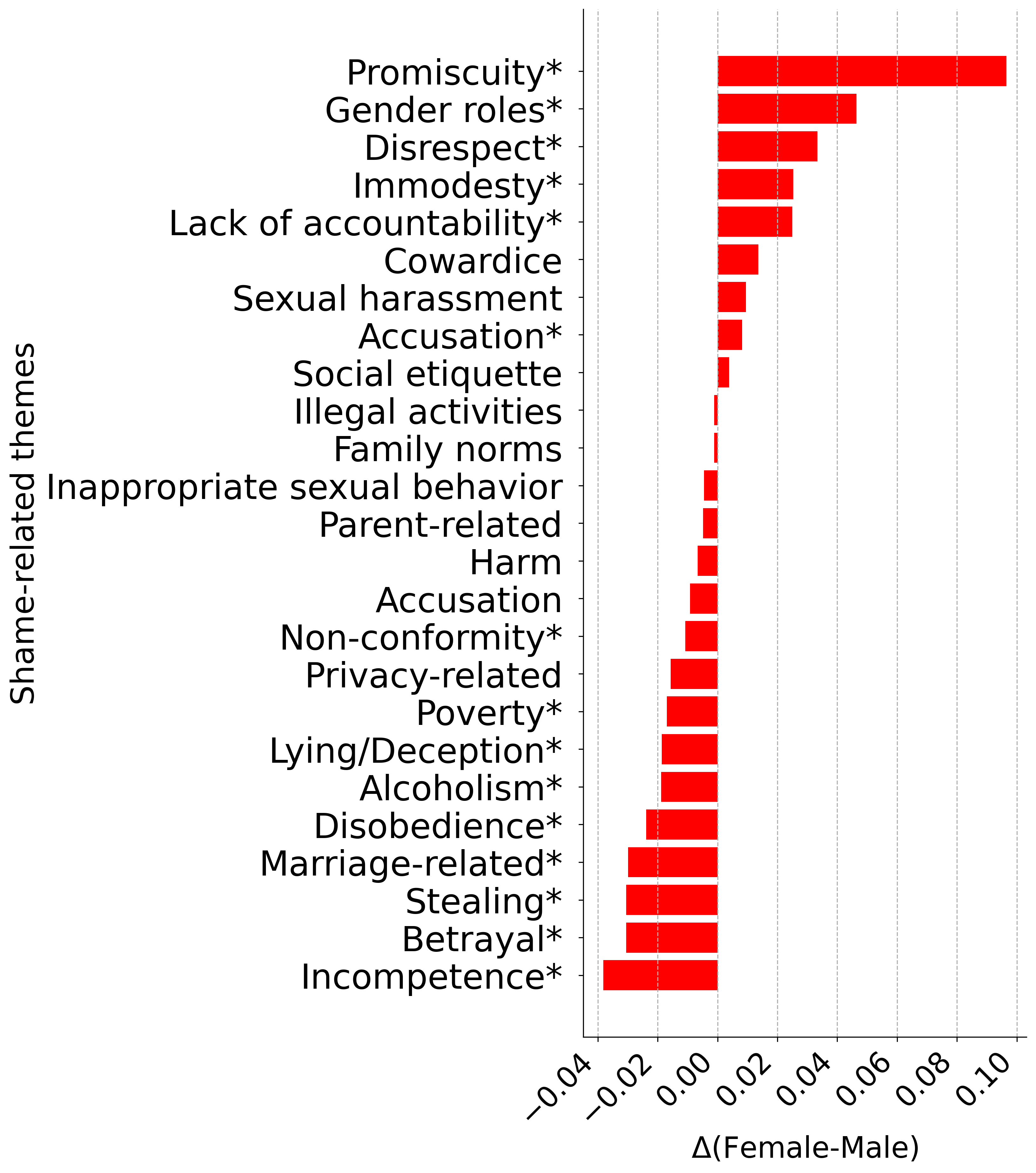}
    \caption{Gender differences in shame-related themes in \textbf{Bollywood} movies. $*$ indicates significant difference i.e., $p<0.05$.}
    \label{fig:shamexgender_bolly}
\end{figure}

\begin{figure}[t]
    \centering
    \includegraphics[width=\linewidth]{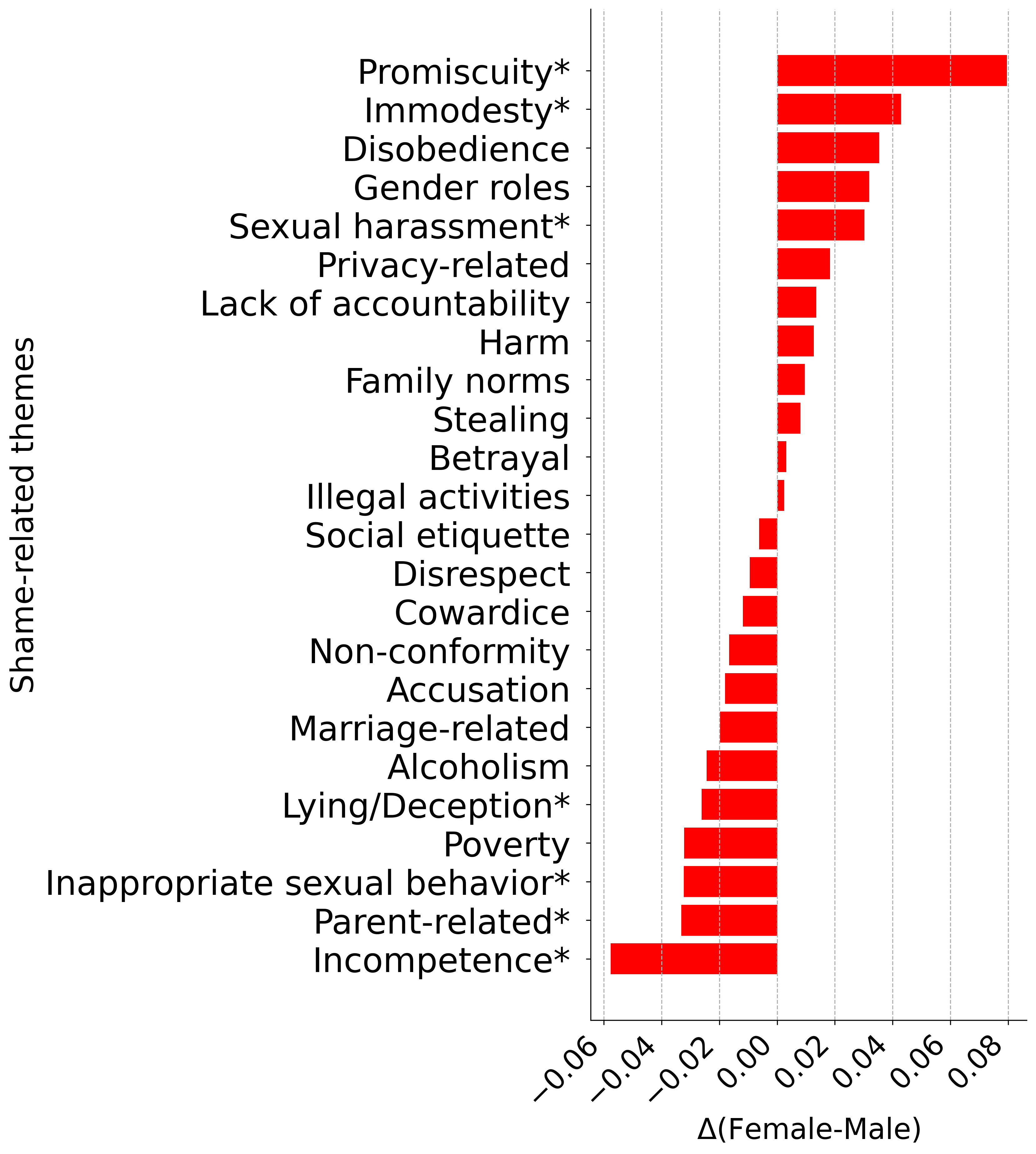}
    \caption{Gender differences in shame-related themes in \textbf{Hollywood} movies. $*$ indicates significant difference i.e., $p<0.05$.}
    \label{fig:shamexgender_holly}
\end{figure}

Social expectations evoking shame are similar across movie industries (See Fig. \ref{fig:shamexgender_bolly} and \ref{fig:shamexgender_holly}). Sexuality (e.g., \textit{promiscuity, immodesty}) is the dominant theme for sanctioning women in both movie industries whereas males are shamed for \textit{incompetency}. Women across movie industries express more pride in \textit{family roles}.  Men-pride in Bollywood is centered on  \textit{justice, winning} and \textit{bravery} whereas it is \textit{duty}, \textit{self-identity} and \textit{winning} in Hollywood (See Fig. \ref{fig:genderxnorms_pride} for gender differences in pride). 

\section{Discussion}

 We release a cross-cultural dataset of shame- and pride-related expressions with implicit reasons behind them. To the best of our knowledge, the only related corpus is the GoEmotions dataset \cite{demszky2020goemotions} comprising 817 samples for embarrassment and 452 samples for pride. Our dataset can help develop computational models for detecting shame/pride expressions and recognizing how appraisals of such emotions vary -- enabling culturally cognisant human-computer conversations for diverse users \cite{kim2011shame} and cross-cultural intent translation\footnote{DARPA Computational Cultural Understanding: \url{https://www.darpa.mil/program/computational-cultural-understanding}}. 
 
Shame is a highly undesirable emotion emphasizing incompetency and failures in the U.S. and used sparingly \cite{cohen2003american, boiger2013emotions} whereas Eastern cultures use shame more frequently and it is known to have a desirable affect \cite{lim2016cultural}. Notably, we also observe that shame is 4.5 times more common in Bollywood compared to Hollywood. We also observe that shame is self-focused (reflecting internal shame) in Hollywood (See Table \ref{tab:data_stats} and Fig. \ref{fig:shame_compare} for frequency distribution and psychosocial constructs). In contrast, shame is interdependent in collectivist communities \cite{wong2007cultural}. The empirical analyses reveal other-focused (also, the public nature of shame) in Bollywood movies (See Fig. \ref{fig:pride_compare}). Moreover, the contrasting tenses coupled with varying emotions, i.e., past-focus + sadness in Hollywood vs. present-focus + conflict in Bollywood, reflect their varying goals, i.e., remorse for past failures/losses vs enforcing conformity \cite{wong2007cultural}.  
Pride-related discourse in Hollywood is duty and achievement-focused, in line with prior findings that the significance of "success" grows as a society becomes more individualistic \cite{cohen2003american}. In Bollywood, pride is collective and male-focused (\textit{we, achievement, male-focused}) \cite{khadilkar2022gender}. It is also interesting to note linguistic markers such as \textit{determiner} (used with objects/nouns) in Bollywood indicate more materialistic pride, compared to personal pronouns and social references in Hollywood which reflect more personal pride.

Eliciting reasons leading to shame and pride expressions revealed social expectations across cultures. We observed high specificity in extracted reasons revealing cultural subtleties (e.g., \textit{desire for a son} in Bollywood vs. \textit{ returning calls after date night} in Hollywood). To tease apart cultural differences, we first considered mapping social expectations/generated reasons to Schwartz's Theory of Values \cite{schwartz2012overview}. However, social expectations were found to have contrasting values depending on the culture. Consider "refusing to marry", an instance of non-conformity in Indian society, whereas it is an instance of self-direction in a Western context. 
 We thus performed hierarchical clustering and empirically picked the distance after manually analyzing the quality and granularity of clusters. The situations connected to shame and pride in the U.S. society (i.e., \textit{lacking accountability, poverty, harm}) vs Indian society (e.g., \textit{inappropriate sexual behavior, gender roles, betrayal} - see Fig \ref{fig:cluster_shame} and \ref{fig:cluster_pride}) align with known cultural tendencies (i.e., individualist vs collectivist) of both nations \cite{triandis1988collectivism}.

 The vision of safe and accountable AI is centered on LLMs' moral and value alignment. 
The dominant approach for norm discovery involves prompting LLMs, sometimes coupled with a verification step such as an entailment test or identifying underlying emotion (negative emotion $\rightarrow$ norm violation) \cite{jiang2021delphi, fung2022normsage, ch2023sociocultural}. 
However, a majority of social situations and the human annotators employed to label those situations reflect English beliefs and ethics. 
Our approach for extracting culture-specific social expectations and attitudes using social emotions overcomes this limitation. While the clustering was performed on the reasons extracted from movie conversations, the cluster themes are similar to patterns seen during LIWC analysis, validating our approach for overcoming Anglocentric bias in LLMs for norm discovery.

Lastly, we examined how social expectations associated with shame/pride are attributed to men and women across cultures influencing their social behaviors. For example, women are less assertive across cultures to avoid negative attribution \cite{amanatullah2010negotiating, ferguson2000engendering} and ask for less during negotiation than men \cite{arnold2021children}. We demonstrated that shame, a negative self-conscious emotion expressing devaluation, is targeted toward women more than men, and pride, a social emotion endorsing social value, is more used for men than women. 
The similarity in social expectations (e.g. regulating sexual behaviors of women vs shaming incompetency in men) across cultures is surprising. We thus feel that it is important to characterize social biases in the training data before their use for aligning LLMs.

\section{Conclusion}
We introduced a multi-cultural dataset of shame/pride-related expressions and the underlying social expectations. This study demonstrates (a) cross-cultural linguistic differences in shame and pride-related expression, offering insights into their functions across cultures, (b) cultural dichotomy in social expectations, and (c) more "social sanctions" and fewer "endorsements" to women in movies. Future work can utilize our dataset for culturally aligned LLMs and build social emotion perception and appraisal in human-computer interactions. Additionally, our analysis of shame/pride is the first of its kind; we hope future NLP researchers will build upon this work to investigate social expectations in LLMs from a multicultural, social emotion-based perspective.

\section*{Social Impact and Ethics Statement}

Social norms discovery is a crucial component of social and behavioral change programs\footnote{ALIGN-\url{https://www.alignplatform.org/learning-collaborative}} promoting equity, social justice, and well-being \cite{mauduy2022combining, bonan2020interaction}. Further work explores style as a product of norm differences \cite{havaldar2023comparing}.
Social psychology investigates social norms (descriptive vs injunctive) to design experiments for understanding behaviors such as self-regulation, persuasion \cite{cialdini1990focus} and decision-making \cite{gavrilets2020dynamics, bhanot2021isolating} to promote collective-level change in societies. 

Relatedly, \citet{kimbrough2023theory} showed people's tendency to choose self-serving social norms using a dictator-recipient setup, emphasizing the need for dedicated research efforts to understand morality and belief distortion in different contexts. The norms and cultural preferences learned from movies that often showcase stereotypical behaviors of society may induce pluralistic ignorance and, more importantly, lead to discrimination and biases in LLMs when used for training. We hope that this paper will encourage scrutiny of source corpora and derived norms before their use for fine-tuning LLMs.

\section*{Limitations}
Social norms mutate as society evolves. We acknowledge that our dataset of movies (released post-1990) may reflect social norms that are less characteristic of contemporary society. Moreover, countries like India and America contain a mix of cultures. The captured norms may not reflect the cultural variations, for example, between regions (e.g., East Coast vs West Coast in the U.S.A or North vs South in India). Movies also exaggeratedly depict the world around us (e.g., weddings, criminal activities, sexual abuse, etc.), and we caution against stereotyping cultures based on movie-based norms. 

The dominant language in Bollywood movies is Hindi and our analysis is based on their English translations which may not always be accurate, especially when the discourse is about 
concepts native to a culture. LIWC dictionary may also lack complete coverage for such concepts.  We did not compare the movie genre and acknowledge that situational/unrealistic norms (e.g., science fiction, comedy, etc.) could exist. Regardless, social norms associated with shame and pride are still relatable and reflect the target audience's beliefs. This study is conducted over a thirty-year long period and the aggregated norms may not truly reflect the current trends.

 We acknowledge that the excerpts expressing shame and pride from movie subtitles may have led to the over-representation of certain social situations (e.g., son's achievement and daughter's wedding in India vs duty and competence in America). The prompts used to elicit norms may have induced unwanted bias \cite{cheng2023compost, lucy2021gender} and it is worth investigating the variations if any, in extracted norms due to different prompt designs.

\section*{Data use and availability}
Bollywood subtitles were scraped from a website called \url{www.bollynook.com} and Hollywood subtitles are taken from publicly available Kaggle movie subtitles datasets.  We will only release dialogue excerpts related to shame and pride. We have around 7.5k such instances in Bollywood and almost 2.8k in Hollywood. Under the interpretation of fair use in research, this makes up a very small portion of dialogues taken from Bollywood (i.e. almost 3 samples (5 lines long) per movie on average) and Hollywood (about 1 sample (5 lines long) per movie on average).

\section*{Acknowledgements}
This work was supported in part by the Penn Global Engagement Research Fund. We acknowledge the use of ChatGPT, an AI language model developed by OpenAI, for its assistance with writing.

\bibliography{acl}

\begin{thebibliography}{49}
\expandafter\ifx\csname natexlab\endcsname\relax\def\natexlab#1{#1}\fi

\bibitem[{Achiam et~al.(2023)Achiam, Adler, Agarwal, Ahmad, Akkaya, Aleman, Almeida, Altenschmidt, Altman, Anadkat et~al.}]{achiam2023gpt}
Josh Achiam, Steven Adler, Sandhini Agarwal, Lama Ahmad, Ilge Akkaya, Florencia~Leoni Aleman, Diogo Almeida, Janko Altenschmidt, Sam Altman, Shyamal Anadkat, et~al. 2023.
\newblock Gpt-4 technical report.
\newblock \emph{arXiv preprint arXiv:2303.08774}.

\bibitem[{Adkins and Castle(2014)}]{adkins2014moving}
Todd Adkins and Jeremiah~J Castle. 2014.
\newblock Moving pictures? experimental evidence of cinematic influence on political attitudes.
\newblock \emph{Social Science Quarterly}, 95(5):1230--1244.

\bibitem[{Amanatullah and Morris(2010)}]{amanatullah2010negotiating}
Emily~T Amanatullah and Michael~W Morris. 2010.
\newblock Negotiating gender roles: Gender differences in assertive negotiating are mediated by women’s fear of backlash and attenuated when negotiating on behalf of others.
\newblock \emph{Journal of personality and social psychology}, 98(2):256.

\bibitem[{Arnold and McAuliffe(2021)}]{arnold2021children}
Sophie~H Arnold and Katherine McAuliffe. 2021.
\newblock Children show a gender gap in negotiation.
\newblock \emph{Psychological Science}, 32(2):153--158.

\bibitem[{Atari et~al.(2023)Atari, Xue, Park, Blasi, and Henrich}]{atari2023humans}
Mohammad Atari, Mona~J Xue, Peter~S Park, Dami{\'a}n Blasi, and Joseph Henrich. 2023.
\newblock Which humans?
\newblock \emph{Working paper}.

\bibitem[{Barnard(1947)}]{barnard1947significance}
GA~Barnard. 1947.
\newblock Significance tests for 2$\times$ 2 tables.
\newblock \emph{Biometrika}, 34(1/2):123--138.

\bibitem[{Bhanot(2021)}]{bhanot2021isolating}
Syon~P Bhanot. 2021.
\newblock Isolating the effect of injunctive norms on conservation behavior: New evidence from a field experiment in california.
\newblock \emph{Organizational Behavior and Human Decision Processes}, 163:30--42.

\bibitem[{Boiger et~al.(2013)Boiger, Deyne, and Mesquita}]{boiger2013emotions}
Michael Boiger, Simon~De Deyne, and Batja Mesquita. 2013.
\newblock Emotions in “the world”: cultural practices, products, and meanings of anger and shame in two individualist cultures.
\newblock \emph{Frontiers in psychology}, 4:867.

\bibitem[{Bonan et~al.(2020)Bonan, Cattaneo, d’Adda, and Tavoni}]{bonan2020interaction}
Jacopo Bonan, Cristina Cattaneo, Giovanna d’Adda, and Massimo Tavoni. 2020.
\newblock The interaction of descriptive and injunctive social norms in promoting energy conservation.
\newblock \emph{Nature Energy}, 5(11):900--909.

\bibitem[{Boyd et~al.(2022)Boyd, Ashokkumar, Seraj, and Pennebaker}]{boyd2022development}
Ryan~L Boyd, Ashwini Ashokkumar, Sarah Seraj, and James~W Pennebaker. 2022.
\newblock The development and psychometric properties of liwc-22.
\newblock \emph{Austin, TX: University of Texas at Austin}, pages 1--47.

\bibitem[{Caffaro et~al.(2014)Caffaro, Ferraris, and Schmidt}]{caffaro2014gender}
Federica Caffaro, Federico Ferraris, and Susanna Schmidt. 2014.
\newblock Gender differences in the perception of honour killing in individualist versus collectivistic cultures: Comparison between italy and turkey.
\newblock \emph{Sex roles}, 71(9):296--318.

\bibitem[{CH-Wang et~al.(2023)CH-Wang, Saakyan, Li, Yu, and Muresan}]{ch2023sociocultural}
Sky CH-Wang, Arkadiy Saakyan, Oliver Li, Zhou Yu, and Smaranda Muresan. 2023.
\newblock Sociocultural norm similarities and differences via situational alignment and explainable textual entailment.
\newblock \emph{arXiv preprint arXiv:2305.14492}.

\bibitem[{Cheng et~al.(2023)Cheng, Piccardi, and Yang}]{cheng2023compost}
Myra Cheng, Tiziano Piccardi, and Diyi Yang. 2023.
\newblock Compost: Characterizing and evaluating caricature in llm simulations.
\newblock \emph{arXiv preprint arXiv:2310.11501}.

\bibitem[{Cialdini et~al.(1990)Cialdini, Reno, and Kallgren}]{cialdini1990focus}
Robert~B Cialdini, Raymond~R Reno, and Carl~A Kallgren. 1990.
\newblock A focus theory of normative conduct: Recycling the concept of norms to reduce littering in public places.
\newblock \emph{Journal of personality and social psychology}, 58(6):1015.

\bibitem[{Cohen(2003)}]{cohen2003american}
Dov Cohen. 2003.
\newblock The american national conversation about (everything but) shame.
\newblock \emph{Social Research: An International Quarterly}, 70(4):1075--1108.

\bibitem[{Cohen et~al.(2011)Cohen, Wolf, Panter, and Insko}]{cohen2011introducing}
Taya~R Cohen, Scott~T Wolf, Abigail~T Panter, and Chester~A Insko. 2011.
\newblock Introducing the gasp scale: a new measure of guilt and shame proneness.
\newblock \emph{Journal of personality and social psychology}, 100(5):947.

\bibitem[{Demszky et~al.(2020)Demszky, Movshovitz-Attias, Ko, Cowen, Nemade, and Ravi}]{demszky2020goemotions}
Dorottya Demszky, Dana Movshovitz-Attias, Jeongwoo Ko, Alan Cowen, Gaurav Nemade, and Sujith Ravi. 2020.
\newblock Goemotions: A dataset of fine-grained emotions.
\newblock \emph{arXiv preprint arXiv:2005.00547}.

\bibitem[{Ferguson and Eyre(2000)}]{ferguson2000engendering}
Tamara~J Ferguson and Heidi~L Eyre. 2000.
\newblock Engendering gender differences in shame and guilt: Stereotypes, socialization, and situational pressures.
\newblock \emph{Gender and emotion}, page 254.

\bibitem[{Fessler(2007)}]{fessler2007appeasement}
D~Fessler. 2007.
\newblock From appeasement to conformity.
\newblock \emph{Self-conscious emotions: Theory and research}, pages 174--193.

\bibitem[{Fischer and Tangney(1995)}]{fischer1995self}
Kurt~W Fischer and June~Price Tangney. 1995.
\newblock Self-conscious emotions and the affect revolution: Framework and overview.
\newblock \emph{Self-conscious emotions: The psychology of shame, guilt, embarrassment, and pride}, pages 3--22.

\bibitem[{Fung et~al.(2022)Fung, Chakraborty, Guo, Rambow, Muresan, and Ji}]{fung2022normsage}
Yi~R Fung, Tuhin Chakraborty, Hao Guo, Owen Rambow, Smaranda Muresan, and Heng Ji. 2022.
\newblock Normsage: Multi-lingual multi-cultural norm discovery from conversations on-the-fly.
\newblock \emph{arXiv preprint arXiv:2210.08604}.

\bibitem[{Gavrilets(2020)}]{gavrilets2020dynamics}
Sergey Gavrilets. 2020.
\newblock The dynamics of injunctive social norms.
\newblock \emph{Evolutionary Human Sciences}, 2:e60.

\bibitem[{Goetz and Keltner(2007)}]{goetz2007shifting}
Jennifer~L Goetz and Dacher Keltner. 2007.
\newblock Shifting meanings of self-conscious emotions across cultures.
\newblock \emph{The self-conscious emotions: Theory and research}, pages 153--173.

\bibitem[{Havaldar et~al.(2023{\natexlab{a}})Havaldar, Pressimone, Wong, and Ungar}]{havaldar2023comparing}
Shreya Havaldar, Matthew Pressimone, Eric Wong, and Lyle Ungar. 2023{\natexlab{a}}.
\newblock \href {https://doi.org/10.18653/v1/2023.emnlp-main.419} {Comparing styles across languages}.
\newblock In \emph{Proceedings of the 2023 Conference on Empirical Methods in Natural Language Processing}, pages 6775--6791, Singapore. Association for Computational Linguistics.

\bibitem[{Havaldar et~al.(2023{\natexlab{b}})Havaldar, Rai, Singhal, Liu, Guntuku, and Ungar}]{havaldarEmotion}
Shreya Havaldar, Sunny Rai, Bhumika Singhal, Langchen Liu, Sharath~Chandra Guntuku, and Lyle Ungar. 2023{\natexlab{b}}.
\newblock Multilingual language models are not multicultural: A case study in emotion.
\newblock In \emph{WASSA'2023, ACL}.

\bibitem[{Jiang et~al.(2021)Jiang, Hwang, Bhagavatula, Bras, Forbes, Borchardt, Liang, Etzioni, Sap, and Choi}]{jiang2021delphi}
Liwei Jiang, Jena~D Hwang, Chandra Bhagavatula, Ronan~Le Bras, Maxwell Forbes, Jon Borchardt, Jenny Liang, Oren Etzioni, Maarten Sap, and Yejin Choi. 2021.
\newblock Delphi: Towards machine ethics and norms.
\newblock \emph{arXiv preprint arXiv:2110.07574}.

\bibitem[{Khadilkar et~al.(2022)Khadilkar, KhudaBukhsh, and Mitchell}]{khadilkar2022gender}
Kunal Khadilkar, Ashiqur~R KhudaBukhsh, and Tom~M Mitchell. 2022.
\newblock Gender bias, social bias, and representation in bollywood and hollywood.
\newblock \emph{Patterns}, 3(2).

\bibitem[{Kim et~al.(2011)Kim, Thibodeau, and Jorgensen}]{kim2011shame}
Sangmoon Kim, Ryan Thibodeau, and Randall~S Jorgensen. 2011.
\newblock Shame, guilt, and depressive symptoms: a meta-analytic review.
\newblock \emph{Psychological bulletin}, 137(1):68.

\bibitem[{Kimbrough and Vostroknutov(2023)}]{kimbrough2023theory}
Erik~O Kimbrough and Alexander Vostroknutov. 2023.
\newblock A theory of injunctive norms.
\newblock \emph{Available at SSRN 3566589}.

\bibitem[{Kubrak(2020)}]{kubrak2020impact}
Tina Kubrak. 2020.
\newblock Impact of films: Changes in young people’s attitudes after watching a movie.
\newblock \emph{Behavioral sciences}, 10(5):86.

\bibitem[{Lewis et~al.(2010)Lewis, Takai-Kawakami, Kawakami, and Sullivan}]{lewis2010cultural}
Michael Lewis, Kiyoko Takai-Kawakami, Kiyobumi Kawakami, and Margaret~Wolan Sullivan. 2010.
\newblock Cultural differences in emotional responses to success and failure.
\newblock \emph{International journal of behavioral development}, 34(1):53--61.

\bibitem[{Lim(2016)}]{lim2016cultural}
Nangyeon Lim. 2016.
\newblock Cultural differences in emotion: differences in emotional arousal level between the east and the west.
\newblock \emph{Integrative medicine research}, 5(2):105--109.

\bibitem[{Lucy and Bamman(2021)}]{lucy2021gender}
Li~Lucy and David Bamman. 2021.
\newblock Gender and representation bias in gpt-3 generated stories.
\newblock In \emph{Proceedings of the Third Workshop on Narrative Understanding}, pages 48--55.

\bibitem[{Mauduy et~al.(2022)Mauduy, Priolo, Margas, and S{\'e}n{\'e}meaud}]{mauduy2022combining}
Maxime Mauduy, Daniel Priolo, Nicolas Margas, and C{\'e}cile S{\'e}n{\'e}meaud. 2022.
\newblock When combining injunctive and descriptive norms strengthens the hypocrisy effect: A test in the field of discrimination.
\newblock \emph{Frontiers in Psychology}, 13:989599.

\bibitem[{Pan et~al.(2023)Pan, Chan, Zou, Li, Basart, Woodside, Zhang, Emmons, and Hendrycks}]{pan2023rewards}
Alexander Pan, Jun~Shern Chan, Andy Zou, Nathaniel Li, Steven Basart, Thomas Woodside, Hanlin Zhang, Scott Emmons, and Dan Hendrycks. 2023.
\newblock Do the rewards justify the means? measuring trade-offs between rewards and ethical behavior in the machiavelli benchmark.
\newblock In \emph{International Conference on Machine Learning}, pages 26837--26867. PMLR.

\bibitem[{Pyszczynski et~al.(2004)Pyszczynski, Greenberg, Solomon, Arndt, and Schimel}]{pyszczynski2004people}
Tom Pyszczynski, Jeff Greenberg, Sheldon Solomon, Jamie Arndt, and Jeff Schimel. 2004.
\newblock Why do people need self-esteem? a theoretical and empirical review.
\newblock \emph{Psychological bulletin}, 130(3):435.

\bibitem[{Reimers and Gurevych(2019)}]{reimers2019sentence}
Nils Reimers and Iryna Gurevych. 2019.
\newblock Sentence-bert: Sentence embeddings using siamese bert-networks.
\newblock In \emph{Proceedings of the 2019 Conference on Empirical Methods in Natural Language Processing and the 9th International Joint Conference on Natural Language Processing (EMNLP-IJCNLP)}, pages 3982--3992.

\bibitem[{Resendiz and Klinger(2023)}]{resendiz2023affective}
Yarik~Menchaca Resendiz and Roman Klinger. 2023.
\newblock Affective natural language generation of event descriptions through fine-grained appraisal conditions.
\newblock \emph{INLG 2023}, page 375.

\bibitem[{Schaumberg and Skowronek(2022)}]{schaumberg2022shame}
Rebecca~L Schaumberg and Samuel~E Skowronek. 2022.
\newblock Shame broadcasts social norms: The positive social effects of shame on norm acquisition and normative behavior.
\newblock \emph{Psychological Science}, 33(8):1257--1277.

\bibitem[{Schwartz(2012)}]{schwartz2012overview}
Shalom~H Schwartz. 2012.
\newblock An overview of the schwartz theory of basic values.
\newblock \emph{Online readings in Psychology and Culture}, 2(1):11.

\bibitem[{Talat et~al.(2021)Talat, Blix, Valvoda, Ganesh, Cotterell, and Williams}]{talat2021word}
Zeerak Talat, Hagen Blix, Josef Valvoda, Maya~Indira Ganesh, Ryan Cotterell, and Adina Williams. 2021.
\newblock A word on machine ethics: A response to jiang et al.(2021).
\newblock \emph{arXiv preprint arXiv:2111.04158}.

\bibitem[{Tangney et~al.(1992)Tangney, Wagner, and Gramzow}]{tangney1992proneness}
June~P Tangney, Patricia Wagner, and Richard Gramzow. 1992.
\newblock Proneness to shame, proneness to guilt, and psychopathology.
\newblock \emph{Journal of abnormal psychology}, 101(3):469.

\bibitem[{T{\"o}rnberg(2023)}]{tornberg2023chatgpt}
Petter T{\"o}rnberg. 2023.
\newblock Chatgpt-4 outperforms experts and crowd workers in annotating political twitter messages with zero-shot learning.
\newblock \emph{arXiv preprint arXiv:2304.06588}.

\bibitem[{Triandis(1988)}]{triandis1988collectivism}
Harry Triandis. 1988.
\newblock Collectivism v. individualism: A reconceptualisation of a basic concept in cross-cultural social psychology.
\newblock In \emph{Cross-cultural studies of personality, attitudes and cognition}, pages 60--95. Springer.

\bibitem[{Triandis(1989)}]{triandis1989self}
Harry~C Triandis. 1989.
\newblock The self and social behavior in differing cultural contexts.
\newblock \emph{Psychological review}, 96(3):506.

\bibitem[{Troiano et~al.(2019)Troiano, Pad{\'o}, and Klinger}]{troiano2019crowdsourcing}
Enrica Troiano, Sebastian Pad{\'o}, and Roman Klinger. 2019.
\newblock Crowdsourcing and validating event-focused emotion corpora for german and english.
\newblock In \emph{Proceedings of the 57th Annual Meeting of the Association for Computational Linguistics}, pages 4005--4011.

\bibitem[{Wegge et~al.(2022)Wegge, Troiano, Oberlaender, and Klinger}]{wegge2022experiencer}
Maximilian Wegge, Enrica Troiano, Laura Ana~Maria Oberlaender, and Roman Klinger. 2022.
\newblock Experiencer-specific emotion and appraisal prediction.
\newblock In \emph{Proceedings of the Fifth Workshop on Natural Language Processing and Computational Social Science (NLP+ CSS)}, pages 25--32.

\bibitem[{Wong and Tsai(2007)}]{wong2007cultural}
Ying Wong and Jeanne Tsai. 2007.
\newblock Cultural models of shame and guilt.
\newblock \emph{The self-conscious emotions: Theory and research}, 209:223.

\bibitem[{Yates(1934)}]{yates1934contingency}
Frank Yates. 1934.
\newblock Contingency tables involving small numbers and the $\chi$ 2 test.
\newblock \emph{Supplement to the Journal of the Royal Statistical Society}, 1(2):217--235.

\end{thebibliography}

\appendix
\onecolumn
\setcounter{figure}{0}
\renewcommand\thefigure{\thesection\arabic{figure}}  
\setcounter{table}{0}
\renewcommand{\thetable}{A\arabic{table}}

\section{Appendix}
\label{sec:appendix}

\subsection{Search Keywords}
Search keywords defined in Table \ref{tab:search_keywords} were used to extract shame- and pride-related dialogues from Bollywood and Hollywood movie subtitles.
\begin{table}[!htb]
    \centering
    \begin{tabular}{c|c}
    Lemma &  Word forms \\ \hline
    \verb|shame| &  shame, shamed, shameful, ashamed \\
  \verb|pride|   & proud, prouder, proudly, pride \\
    \end{tabular}
    \caption{List of search keywords for extracting dialogues.}
    \label{tab:search_keywords}
\end{table}


\subsection{Examining Implicit shame} \label{sec:implicit_shame}

We prompted GPT-4 chat to identify (a) if a given discourse expresses shame, and (b) identify the behavior or action behind shame, for a randomly sampled set of 100 samples without explicit mentions of shame/pride each from Hollywood and Bollywood. We asked two human annotators to evaluate the correctness of GPT-4 chat-generated outputs. See Table \ref{tab:annotation_guidelines} for annotation guidelines.

\subsection{Prompts}

The prompts designed to extract the subject experiencing shame and pride, their gender, and the reason behind the emotions are provided in Tables \ref{tab:prompts_bolly} and \ref{tab:prompts_holly}.

\begin{table*}[ht]
    \centering
    \small
    \begin{tabular}{p{14cm}}

\textbf{Prompts for norm discovery in \textit{Bollywood} movie dialogues} \vspace{1mm} \\ \hline \hline
\vspace{1mm}
 "You are an experienced social scientist and you study Indian society. Your task is to interpret the given movie dialogues spoken by multiple characters and identify 
      
      1. Who is \textbf{{feeling proud?}} Provide the gender. If agent or gender is not explicit, use the pronouns, name, and salutations to guess agent and gender.
      
      2. Identify the reason behind feeling proud. If not explicit, return NA.
     
     The answer should be short and in below CSV format.
     $<$who\_proud, gender, reason
     $>$
      \newline
      Input: Your mother had already given up on me because I was useless hopeless But you were the apple of her eye My child will make me proud look after me He'll support in my old-age But this kid left her all alone.
      
      Output: mother, male, provide care for old parents
      
      Input: I've heard there's a promising young student in your school What's his name? He's made us proud in long jump, we are here to felicitate him Call him Show yourself, Raju Tempre 
      
      Output: authority, NA, Sports achievement
      " \\ \hline
\vspace{1mm}
      "You are an experienced social scientist and you study Indian society. Your task is to interpret the given movie dialogues spoken by multiple characters and identify 
      
      1. Who is \textbf{being shamed}? Predict the gender. If gender is not explicit, use the pronouns, name, and salutations to guess gender.
      
      2. Identify the primary reason for shaming. If not explicit, return NA.

      The answer should be short and in CSV format.
      $<$ who\_shame, gender, reason$>$
      
      Input:  And should we bow before others begging....them to marry our daughters? This shall not happen. Neither will the girls be alive here nor shall....we be ashamed of ourselves. You cannot kill the life which God has given. I won't let you commit the sin. 
      
      Output: girl's parent, NA, not able to marry off their daughters
      
      Input:   Black marketers are now in the open. And the thieves too Politics is in a great mess Shame on this system. There's no democracy Get rid of these politicians The gong has struck..''Our hearts are swaying to it's beats''  
      
      Output: System, NA, poor law and regulations
      " \\ \hline

    \end{tabular}
    \caption{Bollywood: Prompts for norm discovery using GPT-4 Chat. The temperature was set to $0$ to minimize randomness. }
    \label{tab:prompts_bolly}
\end{table*}

\begin{table*}[ht]
    \centering
\small
    \begin{tabular}{p{14cm}}
 
    \textbf{{Prompts for norm discovery in \textit{Hollywood} movie dialogues}} \vspace{1mm} \\ \hline \hline
\vspace{1mm}
   
 "You are an experienced social scientist and you study American society. Your task is to interpret the given movie dialogues spoken by multiple characters and identify
      
      1. Who is \textbf{feeling proud}? Provide the gender. If agent or gender is not explicit, use the pronouns, name, and salutations to guess agent and gender.
      
      2. Identify the reason behind feeling proud? If not explicit, return NA.

      The answer should be short and in below CSV format.
      $<$who\_proud, gender, reason
      $>$
      
      Input: I want to go to Worlds and win gold. I want to go to the 88 Olympics in Seoul and win gold. Good! I'm proud of you. Are you getting the support that you need?  What do you mean sir?
      
      Output: Sir, male, winning olympic gold
      
      Input: Yes. Yes, I did.  I promise, this time I really got the promotion.  - I'm proud of you, son. - Thank you, sir.  Excuse me.  Hi, sweetheart.
      
      Output: father, male, for getting the promotion
      " \\ \hline
\vspace{1mm}
      "You are an experienced social scientist and you study American society. Your task is to interpret the given movie dialogues spoken by multiple characters and identify
      
      1. Who is \textbf{being shamed}? Predict the gender. If gender is not explicit, use the pronouns, name, and salutations to guess gender.
      
      2. Identify the primary reason for shaming. If not explicit, return NA.
      
      The answer should be short and in CSV format.\
      $<$ who\_shame, gender, reason$>$
      
      Input:  You still owe me 100. Remember?  You stiffed Donny for 100 bucks?  Cheapskate. Shame on you.  Pay this man his C-note.  Now I know why they call you the Snake.
      
      Output: NA, male, not returning borrowed money

      Input: You prey on your own people.  You steal from your own people.  Have you no shame!?  - Huh? - Well, we're still here.  Man: Mr. Markopolos, it's all yours.
      
      Output: Snake, male, stealing and preying on people
      " \\ \hline
         
    \end{tabular}
    \caption{Hollywood: Prompts for norm discovery using GPT-4 Chat. The temperature was set to $0$ to minimize randomness.}
    \label{tab:prompts_holly}
\end{table*}
\subsubsection{Annotation}
The annotation guidelines to verify the gender predicted by GPT-4 and the correctness of the reason are provided in Table \ref{tab:annotation_guidelines}. The annotators for Bollywood set and Hollywood set were Indian and American respectively. Both annotators were female, proficient in English language, and well-versed with social norms. During annotation, if the gender or the reason is unclear, the annotators were asked to label "not explicit". We only considered the cases where the gender was predicted to be either male or female. The task is objective and inter-annotator agreement was not computed. The annotators volunteered for the task and were not provided monetary compensation.
\begin{table*}[t]
    \centering
    \begin{tabular}{p{14cm}}
    \\ \toprule
\textbf{Guidelines for Manual Evaluation}
    \\ \midrule
       Step-1. Read the conversation and identify the person feeling ashamed (or being shamed) or proud.\\
       Step-2. Identify the gender. Check gender markers such as Mr/Mrs., s/he, him/her, etc. If the name is provided in the conversation, check if the name is likely to be a male name or female. If not clear, mark "not explicit". \\
       Step-3. Read the reason behind shame/pride. Compare with conversation and determine if the provided reason is the cause for shame/pride. \\ \hline
         
    \end{tabular}
    \caption{Guidelines for Annotation}
    \label{tab:annotation_guidelines}
\end{table*}

\begin{table}[ht]
    \centering
    \begin{tabular}{c|c|c}
    \multicolumn{3}{c}{ Gender Evaluation} \\ \toprule
         & Incorrect & Ambiguous 
         \\ \midrule
     Bollywood    & 8  & 15 (3 Female) 
     \\
     Hollywood &  5& 10 (3 Female) 
     \\ \midrule
     \multicolumn{3}{c}{Social Norms/Reason Evaluation} \\ \bottomrule 
           & Incorrect & Ambiguous  
           \\ \midrule
     Bollywood    & 11  & 9 
     \\
     Hollywood & 2 & 1 
     \\ \midrule
     
    \end{tabular}
    \caption{Manual Evaluation of predicted gender and reasons for randomly sampled 100 samples each from Bollywood and Hollywood.}
    \label{tab:manual_evaluation}
\end{table}

\begin{table}[ht]
    \centering
    \begin{tabular}{p{7cm}}
    \toprule
       \textbf{Bollywood}\\ \midrule
mastering a trick\\
fulfilling father's dreams\\
provide care for old parents\\
fiancee's physical appearance\\
his wealth\\
\midrule
\textbf{Hollywood}\\ \midrule
for being a hard worker regardless of the task\\
being a brilliant student\\
winning Olympic gold\\
achievements and growth\\
coming out as queer\\
\bottomrule

    \end{tabular}
    \caption{A subset of reasons extracted from movie dialogues expressing pride. A total of 3163 unique reasons (Bollywood-1589, Hollywood-1574) were extracted.}
    \label{tab:reason_pride}
\end{table}

\subsection{LIWC Correlation Results}

Tables \ref{tab:liwc_shame_bolly}, \ref{tab:liwc_pride_bolly}, \ref{tab:liwc_shame_holly} and \ref{tab:liwc_pride_holly} contain positively correlated ($p<0.05$) LIWC categories, the most frequent five words for each category, Pearson \textit{r} and confidence interval. 

\begin{table*}[ht]
    \centering
    \small
    \begin{tabular}{l|l|c|c|c}
      LIWC Categories & Top-5 words & \textit{r} & p-value & 95\% CI\\ \toprule
      Negative emotions (EMO\_NEG) & (bad, mad, scared, worry, fear) & 0.330 & 0.000 & [0.315, 0.344] \\  
Negative tone (TONE\_NEG) & (lost, kill, wrong, bad, hit) & 0.249 & 0.000 & [0.233, 0.265] \\ 
POWER & (sir, respect, own, kill, poor) & 0.198 & 0.000 & [0.182, 0.214] \\ 
EMOTION & (love, good, bad, happy, crazy) & 0.168 & 0.000 & [0.152, 0.185] \\ 
 $2^{nd}$ person pronouns (YOU) & (you, your, you're, yourself, you've) & 0.161 & 0.000 & [0.145, 0.178] \\ 
Social references (SOCREFS) & (you, your, he, her, him) & 0.148 & 0.000 & [0.132, 0.165] \\ 
FEELING & (feel, touch, feeling, felt, hard) & 0.133 & 0.000 & [0.117, 0.150] \\ 
DRIVES & (we, our, us, sir, married) & 0.115 & 0.000 & [0.098, 0.132] \\ 
SOCIAL & (you, your, he, her, him) & 0.110 & 0.000 & [0.094, 0.127] \\ 
MORAL & (wrong, innocent, duty, decent, excuse) & 0.102 & 0.000 & [0.085, 0.118] \\ 
AFFECT & (love, good, keep, respect, well) & 0.078 & 0.000 & [0.062, 0.095] \\ 
NEGATE & (not, don't, no, aren't, won't) & 0.065 & 0.000 & [0.049, 0.082] \\ 
FEMALE & (her, she, girl, she's, mom) & 0.060 & 0.000 & [0.043, 0.077] \\ 
Personal Pronouns (PPRON) & (you, i, me, your, my) & 0.058 & 0.000 & [0.041, 0.075] \\ 
FAMILY & (son, married, uncle, dad, mom) & 0.055 & 0.000 & [0.039, 0.072] \\ 
Preposition (PREP) & (to, of, in, for, on) & 0.053 & 0.000 & [0.037, 0.070] \\ 
SEXUAL & (chaste, lust, sex, sexy, pimp) & 0.051 & 0.000 & [0.034, 0.067] \\ 
PRONOUN & (you, i, me, your, my) & 0.040 & 0.000 & [0.023, 0.057] \\ 
Auxiliary verbs (AUXVERB) & (is, are, have, be, don't) & 0.029 & 0.001 & [0.013, 0.046] \\ 
CONFLICT & (kill, killed, accusing, killing, cruel) & 0.024 & 0.007 & [0.007, 0.041] \\ 
SWEAR & (hell, bloody, idiot, damn, ass) & 0.024 & 0.007 & [0.007, 0.041] \\ 
Anger (EMO\_ANGER) & (mad, angry, hate, cruel, argue) & 0.022 & 0.015 & [0.005, 0.039] \\  
FOCUSPRESENT & (is, are, don't, i'm, aren't) & 0.021 & 0.018 & [0.005, 0.038] \\ \bottomrule
      
    \end{tabular}
    \caption{Psychosocial categories positively correlated ($p<0.05$) with \textit{shame} in \textbf{Bollywood}. p-values were corrected using Benjamini-Hochberg correction. The categories are arranged in decreasing order of correlation.}
    \label{tab:liwc_shame_bolly}
\end{table*}

\begin{table*}[ht]
    \centering
    \small
    \begin{tabular}{l|l|c|c|c}
        LIWC Categories & Top-5 words & \textit{r} & p-value & 95\% CI\\ \toprule
DRIVES & (we, our, us, sir, work) & 0.122 & 0.000 & [0.104,0.140] \\ 
 Determiner (DET) & (the, a, my, your, that) & 0.101 & 0.000 & [0.082,0.119] \\ 
ACHIEVE & (work, better, win, best, try) & 0.098 & 0.000 & [0.080,0.116] \\ 
POWER & (sir, own, respect, kill, power) & 0.091 & 0.000 & [0.072,0.109] \\ 
Social References (SOCREFS) & (you, your, he, we, our) & 0.085 & 0.000 & [0.067,0.103] \\ 
Preposition (PREP) & (of, to, in, for, with) & 0.084 & 0.000 & [0.066,0.103] \\ 
MORAL & (wrong, duty, brave, arrogant, useless) & 0.075 & 0.000 & [0.057,0.093] \\ 
Conjunction (CONJ) & (and, but, so, if, as) & 0.075 & 0.000 & [0.056,0.093] \\ 
REWARD & (win, won, glory, success, successful) & 0.071 & 0.000 & [0.052,0.089] \\ 
Positive tone (TONE\_POS) & (love, good, thank, well, great) & 0.064 & 0.000 & [0.046,0.082] \\ 
POLITIC & (nation, army, sultan, president, dynasty) & 0.064 & 0.000 & [0.045,0.082] \\ 
WE & (we, our, us, we'll, let's) & 0.060 & 0.000 & [0.042,0.079] \\ 
SOCIAL & (you, your, he, we, our) & 0.060 & 0.000 & [0.041,0.078] \\ 
FAMILY & (son, papa, married, dad, uncle) & 0.059 & 0.000 & [0.041,0.078] \\ 
AFFILIATION & (we, our, us, dear, we'll) & 0.059 & 0.000 & [0.041,0.077] \\ 
FEELING & (feel, feeling, hard, felt, sense) & 0.055 & 0.000 & [0.036,0.073] \\ 
ETHNICITY & (indian, indians, british, hindi, caste) & 0.054 & 0.000 & [0.036,0.072] \\ 
MALE & (he, his, him, son, sir) & 0.054 & 0.000 & [0.036,0.072] \\ 
CULTURE & (indian, nation, army, car, indians) & 0.044 & 0.000 & [0.026,0.062] \\ 
AFFECT & (love, good, thank, well, great) & 0.040 & 0.000 & [0.021,0.058] \\ 
ARTICLE & (the, a, an, that) & 0.039 & 0.000 & [0.021,0.058] \\ 
Personal Pronouns (PPRON) & (you, i, my, your, me) & 0.036 & 0.000 & [0.017,0.054] \\ 
PROSOCIAL & (thank, please, sorry, respect, gift) & 0.032 & 0.001 & [0.014,0.051] \\ 
FUNCTION & (you, the, i, of, to) & 0.032 & 0.001 & [0.013,0.050] \\ 
$2^{nd}$ person pronouns (YOU) & (you, your, you're, you've, you'll) & 0.030 & 0.003 & [0.011,0.048] \\ 
THEY & (they, their, them, they're, they'll) & 0.029 & 0.003 & [0.011,0.048] \\ 
CERTITUDE & (really, real, surely, proved, actually) & 0.021 & 0.041 & [0.002,0.039] \\ 
Positive Emotion (EMO\_POS) & (love, good, happy, happiness, smile) & 0.020 & 0.049 & [0.001,0.038] \\ \bottomrule
    \end{tabular}
    \caption{Psychosocial categories positively correlated ($p<0.05$) with \textit{pride} in \textbf{Bollywood}. p-values were corrected using Benjamini-Hochberg correction. The categories are arranged in decreasing order of correlation.}
    \label{tab:liwc_pride_bolly}
\end{table*}

\begin{table*}[ht]
    \centering
    \small
    \begin{tabular}{l|l|c|c|c}

       LIWC Categories & Top-5 words & \textit{r} & p & 95\% CI\\ \hline

       Negative emotions (EMO\_NEG) & (sick, pain, fear, bad, afraid) & 0.425 & 0.000 & [0.403, 0.446] \\ 
Negative tone (TONE\_NEG) & (lost, wrong, sick, pain, poor) & 0.355 & 0.000 & [0.331, 0.377] \\ 
EMOTION & (good, love, sick, pain, bad) & 0.290 & 0.000 & [0.266, 0.314] \\ 
POWER & (own, sir, poor, killed, war) & 0.263 & 0.000 & [0.238, 0.287] \\ 
AFFECT & (well, good, love, help, damn) & 0.168 & 0.000 & [0.142, 0.193] \\ 
DRIVES & (we, us, our, work, we're) & 0.152 & 0.000 & [0.127, 0.178] \\ 
FUNCTION & (you, i, the, to, of) & 0.131 & 0.000 & [0.105, 0.157] \\ 
Personal Pronouns (PPRON) & (you, i, me, i'm, my) & 0.111 & 0.000 & [0.085, 0.137] \\ 
MORAL & (wrong, excuse, decent, honest, duty) & 0.110 & 0.000 & [0.084, 0.136] \\ 
$1^{st}$ person sing. pronouns (I) & (i, me, i'm, my, i'll) & 0.102 & 0.000 & [0.076, 0.128] \\ 
Sadness (EMO\_SAD) & (crying, cry, sob, lonely, sad) & 0.096 & 0.000 & [0.070, 0.122] \\ 
NEGATE & (no, not, don't, nothing, never) & 0.090 & 0.000 & [0.063, 0.116] \\ 
Preposition (PREP) & (to, of, in, for, on) & 0.089 & 0.000 & [0.062, 0.115] \\ 
PRONOUN & (you, i, that, it, me) & 0.088 & 0.000 & [0.062, 0.114] \\ 
YOU & (you, your, you're, yourself, you've) & 0.080 & 0.000 & [0.054, 0.106] \\ 
Auxiliary verb (AUXVERB) & (be, i'm, is, was, have) & 0.077 & 0.000 & [0.051, 0.103] \\ 
FOCUSPAST & (was, did, were, been, didn't) & 0.069 & 0.000 & [0.043, 0.096] \\ 
SOCIAL & (you, your, we, he, you're) & 0.060 & 0.000 & [0.033, 0.086] \\ 
Conjunction (CONJ) & (and, so, but, if, when) & 0.056 & 0.000 & [0.030, 0.082] \\ 
LINGUISTIC & (you, i, the, to, of) & 0.056 & 0.000 & [0.030, 0.082] \\ 
ALLNONE & (no, all, nothing, never, yes) & 0.055 & 0.000 & [0.029, 0.081] \\ 
FAMILY & (son, dad, baby, mom, mama) & 0.054 & 0.000 & [0.028, 0.081] \\ 
Social references (SOCREFS) & (you, your, we, he, you're) & 0.050 & 0.000 & [0.024, 0.077] \\ 
Anxiety (EMO\_ANX) & (fear, afraid, worry, terrified, scared) & 0.042 & 0.004 & [0.015, 0.068] \\ 
ILLNESS & (sick, pain, pains, flu, sickly) & 0.038 & 0.010 & [0.012, 0.064] \\ 
FEELING & (feel, felt, pain, feeling, hard) & 0.035 & 0.020 & [0.008, 0.061] \\ 
Anger (EMO\_ANGER) & (hate, hated, mad, angry, hates) & 0.033 & 0.026 & [0.007, 0.059] \\ 
DIFFER & (not, but, if, didn't, or) & 0.033 & 0.027 & [0.007, 0.059] \\ 
Discrepancy (DISCREP) & (should, can, would, can't, want) & 0.032 & 0.029 & [0.006, 0.059] \\ 
COGNITION & (no, not, all, know, but) & 0.032 & 0.033 & [0.005, 0.058]

    \end{tabular}
    \caption{Psychosocial categories positively correlated (p<0.05) with \textit{shame} in \textbf{Hollywood}. p-values were corrected using Benjamini-Hochberg correction. The categories are arranged in decreasing order of correlation. }
    \label{tab:liwc_shame_holly}
\end{table*}

\begin{table*}[ht]
    \centering
    \small
    \begin{tabular}{l|l|c|c|c}
  
LIWC Categories & Top-5 words & \textit{r} & p-value & 95\% CI \\ \hline

Personal Pronouns (PPRON) & (you, i, i'm, me, my) & 0.135 & 0.000 & [0.113,0.158] \\  
Social References (SOCREFS) & (you, your, we, he, you're) & 0.131 & 0.000 & [0.108,0.154] \\ 
FAMILY & (son, dad, baby, mom, mama) & 0.121 & 0.000 & [0.099,0.144] \\ 
Conjunction (CONJ) & (and, so, but, as, if) & 0.114 & 0.000 & [0.091,0.137] \\ 
SOCIAL & (you, your, we, he, you're) & 0.102 & 0.000 & [0.079,0.125] \\ 
FUNCTION & (you, i, the, of, to) & 0.100 & 0.000 & [0.077,0.123] \\ 
$1^{st}$ person sing. pronouns (I) & (i, i'm, me, my, i'll) & 0.095 & 0.000 & [0.072,0.118] \\ 
YOU & (you, your, you're, you've, yourself) & 0.091 & 0.000 & [0.068,0.114] \\ 
MALE & (he, his, him, man, son) & 0.084 & 0.000 & [0.061,0.107] \\ 
DRIVES & (we, our, us, we're, dad) & 0.076 & 0.000 & [0.053,0.099] \\ 
Positive tone (TONE\_POS) & (good, well, thank, great, love) & 0.072 & 0.000 & [0.049,0.095] \\ 
PRONOUN & (you, i, i'm, that, it) & 0.072 & 0.000 & [0.048,0.095] \\ 
Auxiliary Verb (AUXVERB) & (i'm, be, is, was, have) & 0.071 & 0.000 & [0.048,0.094] \\ 
Positive Emotion (EMO\_POS) & (good, love, happy, hope, wonderful) & 0.070 & 0.000 & [0.047,0.093] \\ 
Preposition (PREP) & (of, to, in, for, on) & 0.070 & 0.000 & [0.047,0.093] \\ 
AFFILIATION & (we, our, us, we're, dad) & 0.054 & 0.000 & [0.031,0.078] \\ 
EMOTION & (good, love, happy, hope, bad) & 0.052 & 0.000 & [0.029,0.075] \\ 
ETHNICITY & (american, irish, chinese, german, christian) & 0.052 & 0.000 & [0.029,0.075] \\ 
REWARD & (win, won, winner, successful, earned) & 0.048 & 0.000 & [0.025,0.071] \\ 
ACHIEVE & (work, better, best, trying, try) & 0.045 & 0.000 & [0.022,0.068] \\ 
POWER & (sir, own, war, strong, mighty) & 0.044 & 0.001 & [0.021,0.067] \\ 
AFFECT & (good, well, thank, great, love) & 0.041 & 0.001 & [0.018,0.064] \\ 
MORAL & (wrong, excuse, hero, brave, dignity) & 0.040 & 0.002 & [0.017,0.064] \\ 
FEMALE & (her, she, she's, girl, ladies) & 0.034 & 0.009 & [0.011,0.057] \\ 
CULTURE & (american, car, president, nation, mayor) & 0.033 & 0.012 & [0.009,0.056] \\ 
FOCUSPAST & (was, did, been, were, had) & 0.032 & 0.013 & [0.009,0.055] \\ 
SHEHE & (he, his, him, her, she) & 0.031 & 0.016 & [0.008,0.054] \\ 
WORK & (work, job, school, deal, company) & 0.028 & 0.033 & [0.005,0.051] 

    \end{tabular}
    \caption{Psychosocial categories positively correlated (p<0.05) with \textit{pride} in \textbf{Hollywood}. p-values were corrected using Benjamini-Hochberg correction. The categories are arranged in decreasing order of correlation. }
    \label{tab:liwc_pride_holly}
\end{table*}

\subsection{Clustering Results}

Tables \ref{tab:clusterThemes_shame} and \ref{tab:clusterThemes_pride} contain the manually annotated \textit{Cluster Themes}, the total number of samples in each cluster, and Bollywood vs Hollywood distribution. The distance was set to $5$ and the duplicates were removed. The theme and top three examples demonstrating its meaning are provided in Tables \ref{tab:shame_clusterEx} and \ref{tab:pride_clusterEx}.

\begin{table*}[ht]
    \centering
   \small
   \begin{tabular}{p{3.7cm}|p{10.5cm}}
   \toprule
        Theme & Examples from cluster \\  \midrule
Stealing & {shamelessness and taking money, lying and misusing money, Stealing and bribery} \\ 
Poverty & not having money, being poor, living in poor conditions \\
Incompetence & not fulfilling responsibilities, not meeting expectations or making a mistake, Failure or perceived incompetence \\
Promiscuity & being promiscuous, Having a love affair, shameless behavior and expressing love \\ 
\textit{Inappropriate social behavior} & not living up to someone's expectations, being disloyal, Being rude and ungrateful \\ 
Disobedience & not upholding cultural values, disobedience or lack of respect, being disloyal \\ 
Parent-related & not standing up for his mother, not taking care of his father, not taking responsibility for his son \\ 
Marriage-related & refusing to marry, being forced into marriage, not accepting the proposed marriage \\ 
Immodesty & Wearing inappropriate clothes, inappropriate behavior nudity in public, behaving indecently in public \\ 
Gender Roles & not behaving as per her husband's expectations, behaving inappropriately in front of her daughter, going against her husband \\ 
Sexual Harassment & assaulting a girl, physical assault on women, sexual assault \\ 
Cowardice & lack of courage, cowardice and inability to stand up for oneself, lack of pride and integrity \\ 
Alcoholism & excessive drinking, inappropriate behavior due to alcohol, drinking and irresponsible behavior \\ 
Non-conformity & not conforming to gender norms, breaking gender norms, behaving inappropriately according to societal norms \\ 
Illegal activities & engaging in criminal activities, Being a criminal, committing illegal acts \\ 
Betrayal & Betrayal or dishonesty, betrayal and infidelity, Deception/Betrayal \\ 
\textit{Lack of accountability} & not taking responsibility and blaming others, wrongdoing without remorse, not acknowledging wrongdoing \\ 
Disrespect & disrespecting others, disrespecting an authority figure, Disrespectful and inappropriate behavior \\ 
Harm & causing harm to others, committing harmful deeds, causing trouble and endangering others \\ 
Lying/Deception & lying and hiding information, Lying or deceit, Deception/Not being truthful \\ 
Inappropriate sexual behavior & inappropriate advances and comments, inappropriate behavior towards a young girl, inappropriate language and behavior \\ 
Social Etiquette & Being humiliated in public, being disrespected and belittled, being mocked and treated shamefully \\ 
Privacy-related & invading personal space, invading someone's privacy, intrusion of personal space, trespassing/invading personal space  \\ 
Accusation & accused of wrongdoing, being accused of something, being accused of infidelity \\ 
Family norms & Disrespecting family, causing harm and shame to family, causing difficulty and shame to family \\
\textbf{Lack of shame} & perceived shamelessness, being perceived as shameless, feeling embarrassed and ashamed \\ \bottomrule
   \end{tabular}
    \caption{Themes ($n=26$) and top-3 examples for clusters obtained after agglomerative clustering of shame-related dialogues from Bollywood and Hollywood. The cluster \textit{Lack of shame} was removed since the reason behind shame was not evident. Clusters \textit{Lack of accountability} and \textit{Inappropriate social behavior} were merged together (and named ``Lack of accountability'') due to overlapping normative expectations.}
    \label{tab:shame_clusterEx}
\end{table*}

\begin{table*}[ht]
    \centering
  
   \small
   \begin{tabular}{p{4cm}|p{10.5cm}}
   \toprule
 Theme & Examples from cluster \\ \midrule
Achievement & being a brilliant student, working hard and achieving a spectacular result, professional achievement \\ 
Family Roles & raising their child, having an unselfish mother, having a simple and good daughter \\ 
Self-identity & being a man, being a woman, being gay \\ 
Duty & fulfilling duty, Accomplishing something great, saving the world \\ 
{Bravery}  & bravery and sacrifice, bravery and service to nation, bravery and selflessness \\ 
{Nation} & saving the country's pride, making country proud, Making country proud\\
Doing the "right" thing & for doing the right thing, for being heroes, for having courage \\ 
Son's achievements & son's success and progress, son's hard work and achievement, son's determination \\ 
Family Honor & maintaining dignity and pride, taking care of family pride, saving family's pride \\ 
Justice & standing up for justice, fighting for justice, bringing justice \\ 
Wedding & marrying with pride and respect, sister's marriage, daughter's marriage into a reputed family \\ 
Physical Appearance \& Assets & his manhood, his wealth and beautiful wife, his knowledge and power \\ 
Resilience & resilience and determination, overcoming struggles, enduring hardships without complaint \\
Winning & Sports achievement, working hard and achieving a spectacular result, winning something \\ 
Ethnolinguistic Identity & being Indian, defending the pride of Rajputs, being a Maharashtrian \\ \bottomrule
   \end{tabular}
    \caption{Themes ($n=15$) and top-3 examples for clusters obtained after agglomerative clustering of pride-related dialogues from Bollywood and Hollywood. }
    \label{tab:pride_clusterEx}
\end{table*}

\begin{table*}[ht]
    \centering
    \begin{tabular}{l|c|c|c}
    \toprule
        Theme (Shame-related norms) & Total Samples & Hollywood & Bollywood \\
        \midrule
Lack of Accountability & 802 & 185 & 617 \\
Accusation & 70 & 0 & 70 \\
Alcoholism  & 98 & 18 & 80 \\
Betrayal & 262 & 26 & 236 \\
Cowardice & 209 & 51 & 158 \\
Disobedience & 216 & 49 & 167 \\
Disrespect & 246 & 26 & 220 \\
Family norms & 203 & 24 & 179 \\
Gender roles & 221 & 17 & 204 \\
Harm & 225 & 64 & 161 \\
Illegal activities & 253 & 27 & 226 \\
Immodesty  & 311 & 58 & 253 \\ 
Incompetence & 342 & 83 & 259 \\ 
Lying/Deception & 222 & 53 & 169 \\
Marriage-related  & 179 & 25 & 154 \\
Non-conformity  & 208 & 28 & 180 \\
Parent-related  & 263 & 34 & 229 \\
Poverty & 414 & 108 & 306 \\
Privacy-related  & 102 & 14 & 88 \\
Promiscuity  & 595 & 89 & 506 \\
Inappropriate Sexual behavior & 407 & 33 & 374 \\
Sexual Harassment & 232 & 24 & 208 \\
Social Etiquette & 417 & 97 & 320 \\
Stealing  & 351 & 49 & 302 \\ \bottomrule
Total & 6848 & 1182 & 5666 
    \end{tabular}
    \caption{Distribution of reasons (shame) across manually labeled clusters. A total of twenty-six clusters were generated with distance=5. Duplicates were removed for clustering.  One cluster with generic reasons ( such as phrases "lack of shame") was removed and two clusters with similar reasons (related to accountability) were merged. Finally, 24 clusters were considered. The total is slightly more than numbers in Table \ref{tab:normGenderDistribution} since a reason could be mapped to multiple clusters.}
    \label{tab:clusterThemes_shame}
\end{table*}

\begin{table*}[ht]
    \centering
     \begin{tabular}{l|c|c|c}
    \toprule
    Theme (Pride-related norms) & Total Samples & Hollywood & Bollywood \\ 
    \midrule
Achievement & 332 & 174 & 158 \\
Bravery & 248 & 69 & 179 \\
Doing the "right" thing & 82 & 82 & 0 \\
Duty & 792 & 449 & 343 \\
Ethnolinguistic Identity & 148 & 7 & 141 \\
Family Honor & 263 & 77 & 186 \\
Family Roles  & 356 & 156 & 200 \\
Justice  & 255 & 109 & 146 \\
Nation & 216 & 57 & 159 \\
Physical Appearance  & 106 & 49 & 57 \\
Resilience  & 180 & 68 & 112 \\
Self-identity & 273 & 157 & 116 \\
Son's Achievements & 315 & 87 & 228 \\
Wedding & 120 & 23 & 97 \\
Winning & 411 & 205 & 206 \\ 
\bottomrule
Total &   4097 & 1769 & 2328
        
    \end{tabular}
   
    \caption{Distribution of reasons (pride) across manually labeled clusters. A total of fifteen clusters were generated with distance=5. The total is slightly more than the numbers in Table \ref{tab:normGenderDistribution} since a reason could be
mapped to multiple clusters}
    \label{tab:clusterThemes_pride}
\end{table*}

\begin{figure*}[ht]
    \centering
    \begin{subfigure}{0.45\textwidth}
        \centering
        \includegraphics[width=\textwidth]{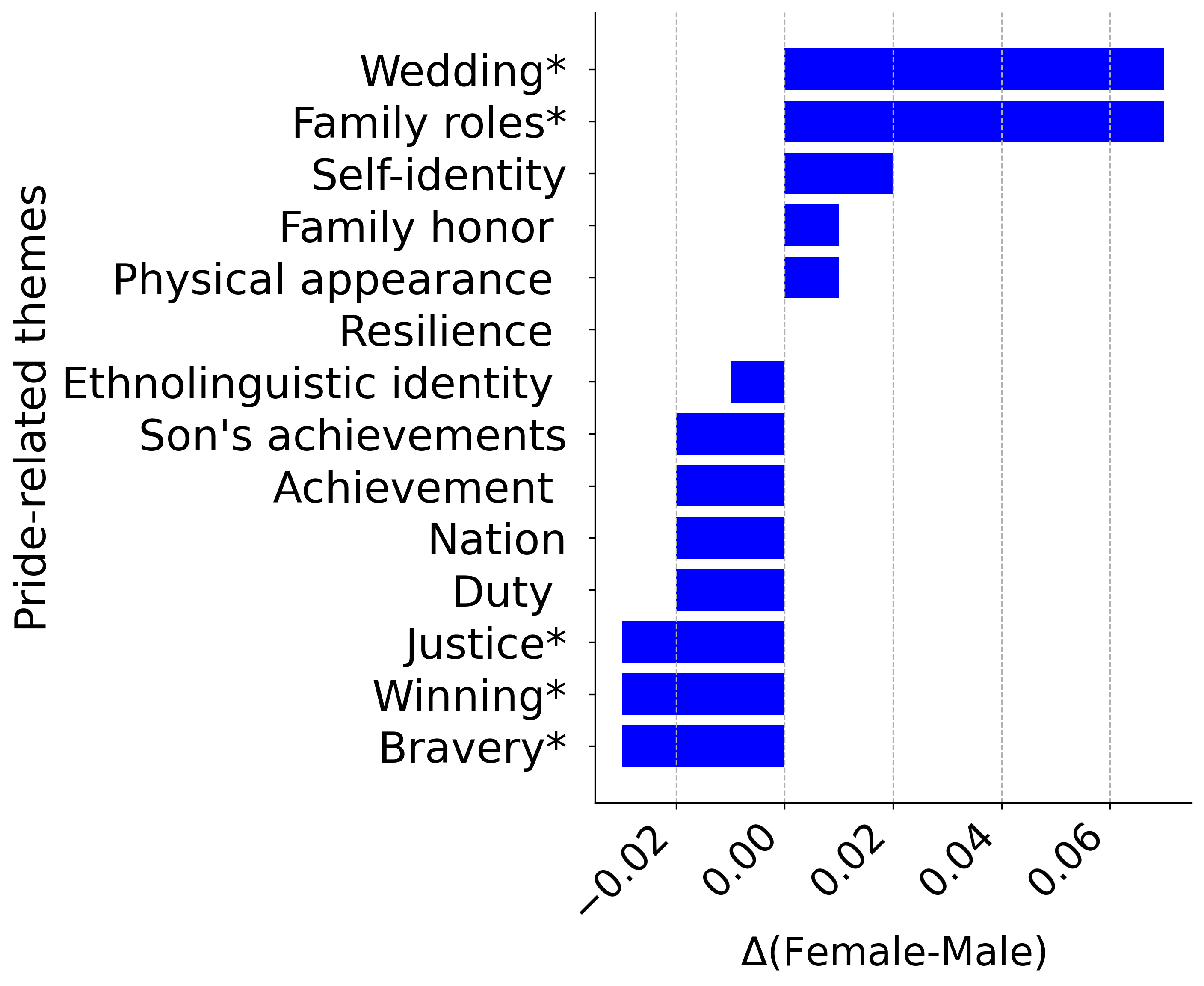}
        \caption{Pride-related norms in \textbf{Bollywood}}
        \label{fig:pride_bolly}
    \end{subfigure}
    \hfill
    \begin{subfigure}{0.45\textwidth}
        \centering
        \includegraphics[width=\textwidth]{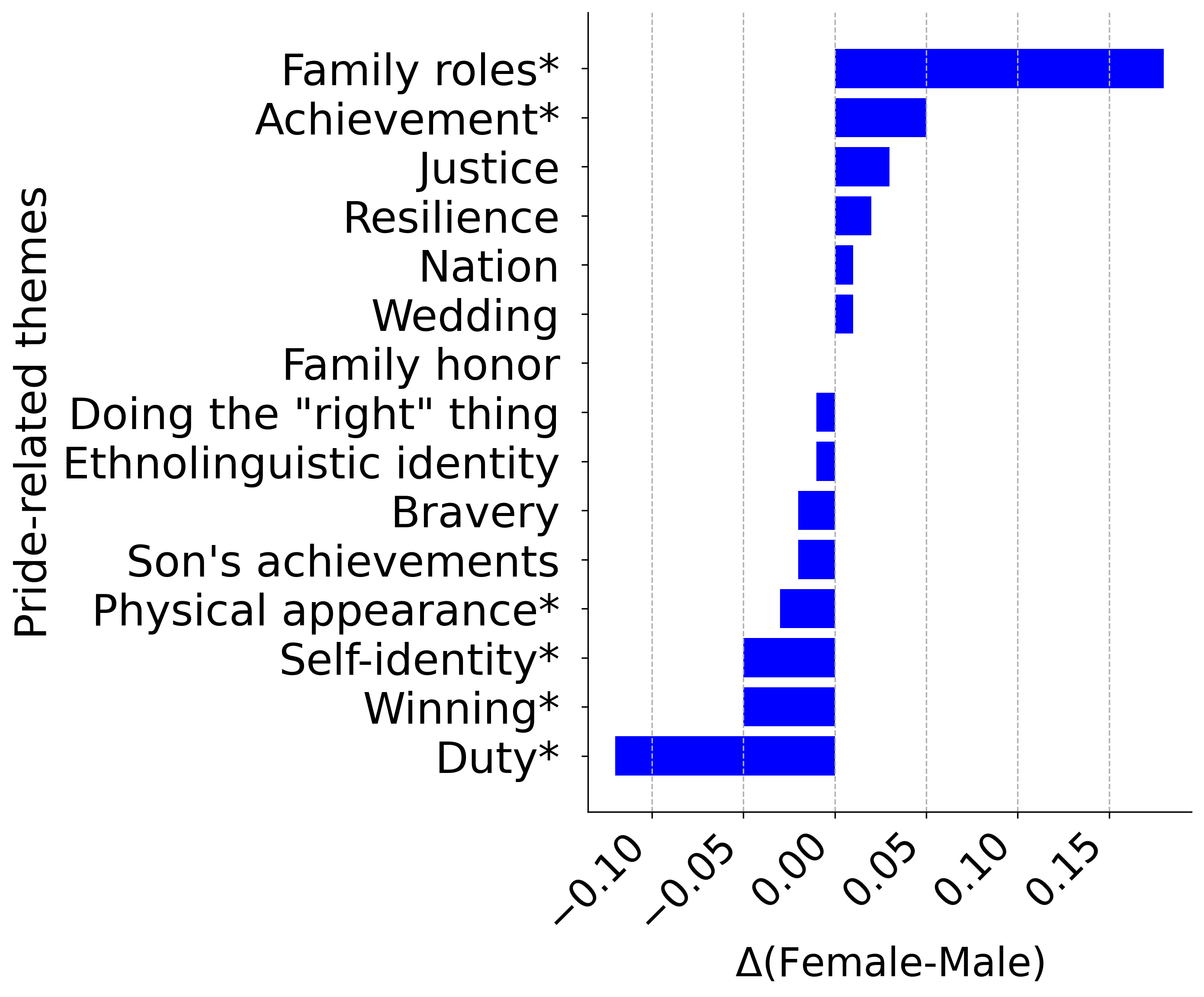}
        \caption{Pride-related norms in\textbf{ Hollywood}}
        \label{fig:pride_holly}
    \end{subfigure}
    \caption{Gender-wise differences in themes associated with normative expectations from pride-related discourse in Bollywood and Hollywood. $\Delta (Female-Male)$ is the difference between normalized dialogues attributed to females under a theme and to males under the same theme (as in eq. \ref{eq:cluster_theme} but for dimension gender). $*$ indicates significant difference i.e., $p<0.05$} 
    \label{fig:genderxnorms_pride}
\end{figure*}

\begin{table}[ht]
    \centering
    \begin{tabular}{l|c |c}
    \toprule
Theme & Statistic & p-value \\ \midrule
Inappropriate Sexual behavior & 5.0380 & 0.00001 \\ 
Gender roles & 3.8262 & 0.0006 \\ 
Betrayal & 3.2044 & 0.001 \\ 
Illegal activities & 2.8258 & 0.005 \\ 
Disrespect & 2.8284 & 0.005 \\ 
Sexual Harassment & 2.8358 & 0.005 \\ 
Promiscuity & 1.5553 & 0.1 \\ 
Accusation & 3.8411 & 0.0006 \\ 
Stealing & 1.6798 & 0.09 \\ 
Parent-related & 1.8961 & 0.06 \\ 
Family norms & 2.0813 & 0.04 \\ 
Non-conformity & 1.4724 & 0.1 \\ 
Marriage-related & 1.1817 & 0.2 \\ 
Privacy-related & 0.9518 & 0.3 \\ 
Alcoholism & -0.2920 & 0.8 \\ 
Immodesty & -0.6634 & 0.5 \\ 
Disobedience & -2.1438 & 0.03 \\ 
Lying/Deception & -2.6508 & 0.008 \\ 
Cowardice & -2.7746 & 0.006 \\ 
Incompetence & -3.5187 & 0.002 \\ 
Social Etiquette & -3.3461 & 0.002 \\ 
Harm & -4.5140 & 0.00009 \\ 
Poverty & -4.9028 & 0.00002 \\ 
Lack of accountability & -4.6312 & 0.00007 \\ \bottomrule
    \end{tabular}
    \caption{Barnard Exact test with Yates correction for testing statistically significant occurrence of shame-related themes in Bollywood and Hollywood. }
    \label{tab:shame_statTest}
\end{table}

\begin{table}[ht]
    \centering
    \begin{tabular}{l|c|c}
    \toprule
       Theme & Statistic & p-value\\ \midrule
Ethnolinguistic Identity & 9.6185 & 0.0000\\ 
Son's Achievements & 5.8025 & 0.0000\\ 
Bravery & 5.0368 & 0.0000\\ 
Family Honor & 4.7046 & 0.0000\\ 
Nation & 5.1183 & 0.0000\\ 
Wedding & 5.3898 & 0.0000\\ 
Resilience & 1.4959 & 0.1370\\ 
Justice & 0.1441 & 0.8963\\ 
Family Roles & -0.2560 & 0.8024\\ 
Physical Appearance & -0.6420 & 0.5223\\ 
Winning & -2.8913 & 0.0038\\ 
Achievement & -3.5425 & 0.0004\\ 
Self-identity & -4.9482 & 0.0000\\ 
Doing the "right" thing & -10.4936 & 0.0000\\ 
Duty & -8.5488 & 0.0000 \\ \bottomrule
    \end{tabular}
    \caption{Barnard Exact test with Yates correction for testing statistically significant occurrence of pride-related themes in Bollywood and Hollywood. }
    \label{tab:pride_statTest}
\end{table}

\end{document}